\documentclass[twocolumn]{aastex631}%
\usepackage{amsmath}	
\usepackage{amssymb}	
\usepackage{rotating}
\usepackage{tikz}
\usepackage{fix-cm}

\newcommand\aastex{AAS\TeX}
\newcommand\latex{La\TeX}

\usepackage{comment}
\usepackage{color,ulem}  
\definecolor{deepblue}{rgb}{0,0,0.5}  
\definecolor{deepred}{rgb}{0.6,0,0}   
\definecolor{deepgreen}{rgb}{0,0.5,0} 
\definecolor{darkgreen}{rgb}{0,0.6,0} 

\def\aliii{{Al\sc{iii}}$\lambda$1860\/}
\def\aliiio{{Al\sc{iii}}\/}
\def\cmed{{$c_{(1/2)}$}\/}

\def\ciiuv{{C\sc{ii}}$\lambda$1335\/}
\def\ciii{{C\sc{iii}]}$\lambda$1909\/}
\def\ciiio{{C\sc{iii}]}\/}
\def\civ{{C\sc{iv}$\lambda$1549}\/}
\def\civo{{C\sc{iv}}\/}
\def\d{$\delta$}

\def\D{$\Delta$\/}

\def\ergs{erg s$^{-1}$}
\def\feii{{Fe\sc{ii}}\/}
\def\feiii{{Fe\sc{iii}}\/}
\def\feiiuv{{Fe\sc{ii}$\lambda$1785}\/}

\def\hb{{\sc{H}}$\beta$\/}
\def\hbbc{{\sc{H}}$\beta_{\rm BC}$}

\def\heii{{He\sc{ii}$\lambda$1640}\/}
\def\heiiopt{{He\sc{ii}$\lambda$4687}\/}
\def\hg{{\sc{H}}$\gamma$\/}
\def\kms{km~s$^{-1}$\/}
\def\l{$\lambda$}

\def\lbol{L$_{\rm bol}$\/}

\def\ledd{L$_{\rm Edd}$\/}

\def\lya{Ly$\alpha$}

\def\nh{n$_{\rm{H}}$}

\def\niii{{N\sc{iii}}$\lambda$1750\/}

\def\niv{{N\sc{iv}}$\lambda$1684\/}
\def\nv{{N\sc{v}}$\lambda$1240\/}

\def\MB{M$_{\rm B}$}
\def\mB{$m_{\rm B}$}
\def\mbh{M$_{\rm BH}$}
\def\mgii{Mg{\sc ii}$\lambda$2800}
\def\mgiio{Mg{\sc ii}}

\def\oiuv{{O\sc{ii}$\lambda$1304}\/}

\def\oiii{{[O\sc{iii}]}\/}

\def\oiiill{{[O\sc{iii}]}$\lambda\lambda$4959,5007\/}

\def\redd{R$_{\rm{Edd}}$\/}
\def\rfe{R$_{\rm{FeII}}$\/}
\def\rmgii{R$_{\rm{MgII}}$\/}

\def\siiiuv{{S\sc{ii}}$\lambda$1263\/}

\def\siii{{Si\sc{ii}$\lambda$1816}\/}
\def\siiii{{Si\sc{iii}]}$\lambda$1892\/}
\def\siiiio{{Si\sc{iii}]}\/}

\def\siiv{{Si\sc{iv}}$\lambda$1397\/}
\def\siivo{{Si\sc{iv}}\/}
\def\s{$\sigma$}
\def\t{$\theta$}

\begin{document}

\title{The Lockman-SpReSO project. Spectroscopic analysis of Type 1 AGN}

\author[0000-0002-1656-827X]{Castalia Alenka Negrete}
\affiliation{Instituto de Astronom\'ia, Universidad Nacional Aut\'onoma de M\'exico, \\
A.P. 70-264, Ciudad de M\'exico, CDMX 04510, Mexico}
\altaffiliation{SECIHTI Research fellow}

\author[0000-0002-9790-6313]{H\'ector J. Ibarra-Medel}
\affiliation{Instituto de Astronom\'ia, Universidad Nacional Aut\'onoma de M\'exico, \\
A.P. 70-264, Ciudad de M\'exico, CDMX 04510, Mexico}


\author[0000-0003-1018-2613]{Erika Ben\'itez}
\affiliation{Instituto de Astronom\'ia, Universidad Nacional Aut\'onoma de M\'exico, \\
A.P. 70-264, Ciudad de M\'exico, CDMX 04510, Mexico}

\author[0000-0002-2653-1120]{Irene Cruz-Gonz\'alez}
\affiliation{Instituto de Astronom\'ia, Universidad Nacional Aut\'onoma de M\'exico, \\
A.P. 70-264, Ciudad de M\'exico, CDMX 04510, Mexico}

\author[0000-0001-6291-5239]{Yair Krongold}
\affiliation{Instituto de Astronom\'ia, Universidad Nacional Aut\'onoma de M\'exico, \\
A.P. 70-264, Ciudad de M\'exico, CDMX 04510, Mexico}

\author{J.Jes\'us Gonz\'alez}
\affiliation{Instituto de Astronom\'ia, Universidad Nacional Aut\'onoma de M\'exico, \\
A.P. 70-264, Ciudad de M\'exico, CDMX 04510, Mexico}

\author{Jordi Cepa}
\affiliation{Instituto de Astrof\'isica de Canarias, E-38205 La Laguna, Tenerife, Spain}
\affiliation{Departamento de Astrof\'isica, Universidad de La Laguna (ULL), E-38205 La Laguna, Tenerife, Spain}

\author{Carmen Padilla-Torres}
\affiliation{Instituto de Astrof\'isica de Canarias, E-38205 La Laguna, Tenerife, Spain}
\affiliation{Departamento de Astrof\'isica, Universidad de La Laguna (ULL), E-38205 La Laguna, Tenerife, Spain}
\affiliation{Fundaci\'on Galileo Galilei - INAF, Rambla Jos\'e Ana Fern\'andez P\'erez, 7, 38712 Bre\~na Baja, TF - Spain}

\author{Miguel Cervi\~no}
\affiliation{Centro de Astrobiolog\'ia (CAB, CSIC-INTA), 28692 ESAC Campus, Villanueva de la Ca\~nada, Madrid, Spain}

\author{Mirjana Povic}
\affiliation{Space Science and Geospatial Institute (SSGI), Entoto Observatory and Research Center (EORC), \\Astronomy and Astrophysics Research Division, PO Box 33679, Addis Abbaba, Ethiopia}
\affiliation{Instituto de Astrof\'isica de Andaluc\'ia (CSIC), 18080 Granada, Spain}
\affiliation{Physics Department, Mbarara University of Science and Technology (MUST), Mbarara, Uganda}

\author{Mart\'in Herrera-Endoqui}
\affiliation{Estancia Posdoctoral por M\'exico, CONAHCYT, Coordinaci\'on de Apoyos a Becarios e Investigadores}
\affiliation{Instituto de Astronom\'ia, Universidad Nacional Aut\'onoma de M\'exico, \\
A.P. 70-264, Ciudad de M\'exico, CDMX 04510, Mexico}

\author{Nancy Jenaro-Ballesteros}
\affiliation{Instituto de Astronom\'ia, Universidad Nacional Aut\'onoma de M\'exico, \\
A.P. 70-264, Ciudad de M\'exico, CDMX 04510, Mexico}

\author{Takamitsu Miyaji}
\affiliation{Instituto de Astronom\'ia -Ensenada, Universidad Nacional Aut\'onona de M\'exico, \\Km. 107 Carret. Tijuana-Ensenada, Ensenada 22860, Mexico}

\author{Mauricio Elias-Ch\'avez}
\affiliation{Instituto de Astronom\'ia -Ensenada, Universidad Nacional Aut\'onona de M\'exico, \\Km. 107 Carret. Tijuana-Ensenada, Ensenada 22860, Mexico}

\author{Miguel S\'anchez-Portal}
\affiliation{Institut de Radioastronomie Millim\'etrique,  Avenida Divina Pastora, 7, Local 20, E 18012 Granada, Spain}

\author{Bernab\'e Cedr\'es}
\affiliation{Instituto de Astrof\'isica de Canarias, E-38205 La Laguna, Tenerife, Spain}

\author{Jacub Nadolny}
\affiliation{Instituto de Astrof\'isica de Canarias, E-38205 La Laguna, Tenerife, Spain}
\affiliation{Astronomical Observatory Institute, Faculty of Physics and Astronomy, \\Adam Mickiewicz University, ul. Sloneczna 36, 60-286 Pozna\'n, Poland} 

\author{Mauro Gonz\'alez-Otero }
\affiliation{Asociaci\'on Astrofísica para la Promoci\'on de la Investigaci\'on,\\ Instrumentaci\'on y su Desarrollo, ASPID, 38205 La Laguna, Tenerife, Spain}

\author{Bereket Assefa}
\affiliation{Space Science and Geospatial Institute (SSGI), Entoto Observatory and Research Center (EORC), \\Astronomy and Astrophysics Research Division, PO Box 33679, Addis Abbaba, Ethiopia}
\affiliation{Debre Berhan University, Debre Berhan, Ethiopia}

\author{H\'ector Hern\'andez-Toledo}
\affiliation{Instituto de Astronom\'ia, Universidad Nacional Aut\'onoma de M\'exico, \\
A.P. 70-264, Ciudad de M\'exico, CDMX 04510, Mexico}

\author{J. Antonio de Diego}
\affiliation{Instituto de Astronom\'ia, Universidad Nacional Aut\'onoma de M\'exico, \\
A.P. 70-264, Ciudad de M\'exico, CDMX 04510, Mexico}

\author{J. Ignacio Gonz\'alez-Serrano}
\affiliation{Dpto. de F\'isica Moderna, Universidad de Cantabria, IFCA, 39005 Santander (Spain)}

\author{A. M. Perez Garcia}
\affiliation{Asociaci\'on Astrofísica para la Promoci\'on de la Investigaci\'on,\\ Instrumentaci\'on y su Desarrollo, ASPID, 38205 La Laguna, Tenerife, Spain}
\affiliation{Centro de Astrobiolog\'ia (CSIC/INTA), 28692 ESAC Campus, Villanueva de la Ca\~nada, Madrid, (Spain)}




\begin{abstract}

We present the first optical-UV spectral systematic analysis of 30 Type 1 AGN selected in the FIR and X-ray in the Lockman-SpReSO Survey. The sample of faint objects ($m_B$=19.6-21.8) covers a large redshift range of 0.33 $> z >$ 4.97 with high S/N ($\sim$21 on average). A detailed spectral analysis based on the Quasar Main Sequence phenomenology prescription was applied to deblend the principal optical-UV emitting regions. Our sample span a bolometric luminosity range of 44.85 $<$ log\lbol\ $<$ 47.87, absolute B-magnitude -20.46 $> M_B >$ -26.14, BH mass of 7.59 $<$ log\mbh\ $<$ 9.80, and Eddington ratio -1.70 $<$ log\redd $<$ 0.56. The analysis shows that 18 high-$z$ objects correspond to Population B, whereas three low-$z$ fall in Populations A2, B1, and B1+. The remaining eight are candidates to be Pop. B and one Pop. A object. None of them are extreme accretors. We looked for tendencies in our sample and compared them with other samples with different selection criteria. 
Evidence for winds was explored using \civ\ line half-height centroid \cmed\ finding wind velocities between 941 and -1587 \kms. This result is consistent with samples with similar ranges of $z$ and $M_B$. The Baldwin effect showed a slope of -0.23 $\pm$ 0.03 dex consistent with previous studies. Spectra from twelve objects in our sample were found in the Sloan Digital Sky Survey Data Release 17 database. We applied the same methodology to compare them to our spectra, finding no evidence of variability.

\end{abstract}



\section{Introduction} \label{sec:intro}
\defcitealias{Gonzalez-Otero2023}{GO23}
\defcitealias{Gonzalez-Otero2024}{GO24}

Extragalactic surveys are fundamental building blocks for studying galaxy evolution using data over a wide range of wavelengths. One of the deep fields largely studied is the Lockman Hole (LH). The LH is the best Galactic window, well known for having small amounts of neutral hydrogen column density ($N_\mathrm{H}$). Its central region has a hydrogen column density of $N_\mathrm{H}=5.8\times 10^{19}\mathrm{cm}^{-2}$ \citep{Lockman1986,Dickey1990}, and has been studied with a wide wavelength coverage \citep[e.g.,][]{Fotopoulou2012, Kondapally2021, Gonzalez-Otero2023}, making it an ideal region for cosmological studies, particularly in the infrared (IR).

The Lockman Spectroscopic Redshift Survey using OSIRIS (Optical System for Imaging and low-Intermediate-Resolution Integrated Spectroscopy) at the Gran Telescopio Canarias \citep[GTC; Lockman-SpReSO;][hereafter \citetalias{Gonzalez-Otero2023}]{Gonzalez-Otero2023}, is a large Guaranteed Time program that provides deep optical spectroscopy of the far-infrared (FIR) sources from the \textit{Herschel}-PEP survey observed at 100 and 160 $\mu$m \citep{Lutz2011}, having optical counterparts in images from OSIRIS in the Sloan Digital Sky Survey (SDSS)-\textit{r} band. 

This survey is intended to complement previous data obtained by space telescopes such as XMM-Newton, Spitzer, and Herschel, as well as radio data.
The Lockman-SpReSO (LS) Survey provided a total of 1144 sources in the central 24\arcmin $\times$ 24\arcmin region of the LH for a limiting magnitude of $R_C \leq$ 24.5 (c.f., Figure 2; \citetalias{Gonzalez-Otero2023}); of these, 107 are stellar, and 1037 are extragalactic sources. Details of the characterization of the LS objects are presented in \citetalias{Gonzalez-Otero2023}. 
In particular, the LS Survey aims to establish the principal properties of active and star-forming galaxies in this field and identify new sources using the obtained spectral data.

Within the extragalactic sources, 114 galaxies with nuclear activity have been identified. To achieve a comprehensive study on Active Galactic Nuclei (AGN), it is important to consider the effects of extinction in the spectra when performing those studies. While a dust extinction correction can be applied, it is always assumed based on a dust model. Therefore, it is important to look for places in the galaxy where the transparency is high enough to allow us to obtain observations where the extinction is minimal. In that sense, the LH is the perfect location to perform and study different types of AGN.

For more than three decades, there have been considerable efforts to organize quasars\footnote{In this work, we use the term quasar and Type 1 AGN equivalently, independent of their luminosity.} based on their observable properties at different wavelengths \citep[e.g.][]{borosongreen92,sulentic00b,shenho14}. The so-called ``Quasar Main Sequence'' \citep[QMS; see][for a recent review]{Panda2024} has evolved from the four-dimensional Eigenvector 1 (4DE1), which comprises optical data including measurements of virialized components of the central region: the FWHM of the \hb\ broad component (\hbbc) and \rfe, the ratio of equivalent widths (EW) of \feii\ (computed in the wavelength range from 4435 to 4685\AA) and \hbbc; data in the UV considering the centroid at half intensity \cmed\ of \civ\ as a measure of the contribution of the non-virialized component due to outflows; and the soft X-ray spectral index as a measure of the contribution of ionizing photons that illuminates the BLR. 
Type 1 AGN have been successfully characterized with the QMS formalism, including high-z objects, as was found recently using the James Webb Space Telescope (JWST) observations \citep[e.g.][]{Loiacono2024}. The QMS finds a dichotomy of Type 1 AGN, dividing them into Population A and B using the FWHM(\hb) = 4000 \kms as a boundary. For instance, evidence supporting this division is the change of the \hb\ profile from Lorentzian (more adequate to fit Pop. A emission line profiles) to double Gaussian, with a redward asymmetry for Pop. B objects  \citep[see][]{Marziani2018}.

This work aims to characterize the Type 1 AGN found in the Lockman-SpReSo survey based on the QMS formalism, and to analyze their properties with other quasar samples at high and low redshifts. A previous study with observations from almost three decades ago reported 43 quasar spectra in the LH \citep[][see also \citealt{Schneider1998} for a single quasar at z=4.45]{Lehmann2000}. 
Other previous works in the optical range of AGN in the LH field have been focused on its photometric properties \citep[e.g.][]{Rovilos2011}.

In addition, studies using AGN in the Lockman-SpReSo project will be addressed in different papers. For instance, the results obtained from the Type 2 AGN sample will be presented in Assefa et al. (in prep.)  along with new identifications of Compton-thick AGN. A multicomponent Spectral Energy Distribution analysis using CIGALE \citep{Boquien2019} will be presented in Herrera-Endoqui et al. (in prep.). Also, taking advantage of deep XMM-Newton data, X-ray spectral analysis will be given for all sources with sufficient counts within Lockman-SpReSO, including AGN (Elias et al. in prep.).

The paper is organized as follows: Section \ref{sec:LHobs} presents the observations and sample selection. In Section \ref{sec:fit}, spectral analysis and fitting details are given. Section \ref{sec:lineComps_AGNParams} shows our AGN derived parmeters, and the analysis is presented in \ref{sec:ana}. In Section \ref{sec:summ}, a summary listing our main results is provided. Throughout this paper, we use the concordance Cosmology with a Hubble constant $H_0=70\,$km\,s$^{-1}$\,Mpc$^{-1}$, $\Omega_m=0.3$, and $\Omega_\Lambda=0.7$.

\section{Lockman-SpReSO observations}
\label{sec:LHobs}

The details of the Lockman-SpReSO observations are described in Sec. 4 of the survey presentation \citetalias{Gonzalez-Otero2023}. In summary, the spectroscopic observations of the faint subset (sources with 20 $\leq R_C \leq$  24.5) were carried out using the OSIRIS instrument at the GTC telescope in MOS-mode, 
hereafter, OSIRIS/GTC. Observations were obtained from 2014 to 2018. The blue region of the spectrum was observed with the R500B grism, which provides a wavelength coverage of 3600-7200 \AA \, and a nominal dispersion of 3.54 \AA\ pixel$^{-1}$. The red part was covered with two grisms (R500R and R1000R). The former has a wavelength coverage of 4800–10000 \AA\ and a dispersion of 4.88 \AA\ pixel$^{-1}$, while the latter has a range of 5100–10000 \AA\ and a dispersion of 2.62 \AA\ pixel$^{-1}$. Table 2 of \citetalias{Gonzalez-Otero2023} provides detailed information about the configuration of the OSIRIS/GTC runs obtained from 2014 to 2018. The data reduction is also described in \S 5 of \citetalias{Gonzalez-Otero2023}. Calibrated 1D spectra were used to obtain redshift estimations and detailed information about each object from the spectral analysis. In this work, we selected the OSIRIS/GTC spectra of a sample of Type 1 AGN identified in the Lockman-SpReSO survey catalog.



\subsection{Sample selection}
\label{sec:selection}

From the initial spectroscopic sample of 409 FIR-selected objects from the Lockman-SpReSO catalog, 69 (17\%) AGN were identified in \citet{Gonzalez-Otero2024}, hereafter, \citetalias{Gonzalez-Otero2024}. The selection criteria are based on X-ray, FIR, and spectroscopic properties, including the use of diagnostic diagrams given by \citet[][]{Baldwin1981_BPT}. 
All the AGN spectra were visually inspected to select only those showing broad emission lines (BELs), i.e., only Type 1 with FWHM larger than 1000 \kms. 
We obtained a final sample of 30 Type 1 AGN in a large redshift interval of 0.46 $<\,z\,<$ 4.97. 

\begin{figure}
    \centering
    \includegraphics[width=1.07\linewidth]{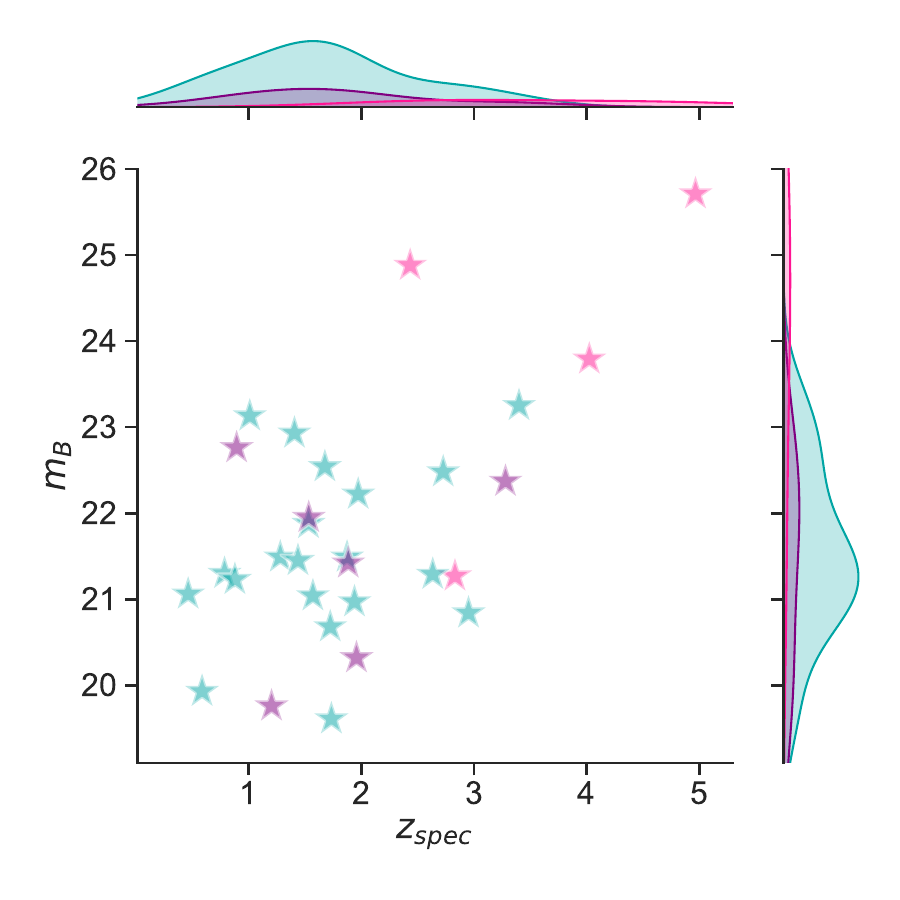}
    \caption{Redshift distribution of  Type 1 AGN in the Lockman-SpReSO as a function of the apparent magnitude. The color code corresponds to the Class Identification of \citetalias{Gonzalez-Otero2023}: FIR in pink, FIRXr in purple, and Xr in blue.}
    \label{fig:zHisto}
\end{figure}

Table~\ref{tab:sample} contains the list of Type 1 AGN considered in this study. \citetalias{Gonzalez-Otero2023} sub-classified 26 of the Type 1 AGN as X-ray point sources (Xr), 10 as FIR objects, one as a red quasar, and one as a sub-millimetric galaxy (SMG). Six objects have both FIR and Xr classifications.
Object 206653 is a BAL (broad absorption line) quasar. All objects were cross-correlated with the SDSS data release 17 \citep[SDSS DR17,][]{Abdurrouf2022} database, and 12 objects with spectroscopic observations were identified. 
The redshift distribution of our sample is presented in Figure \ref{fig:zHisto}, where we can see a broad $z$ distribution with a maximum of objects around $z$ $\sim$ 1.8, nine objects with $z \,>\,$ 2.4 and only three low z objects below 0.8. The apparent magnitude in the B band, \mB, is also plotted to show the faintness of our sample. We can see a tendency of the FIR objects towards high-z high-\mB\ values, while the FIRXr and Xr objects are mixed in a lower-z lower-\mB\ range. 



\begin{table*}
    \centering
    \begin{tabular}{lcccccclc}
\hline
ObjID & RA & DEC & z$_{\rm spec}$ & D$_{\rm L}$ & \MB & GO23 & Fitted lines & SDSS DR17 \\
& deg & deg & & Mpc & mag & class &  & \\
(1) & (2) & (3) & (4) & (5) & (6) & (7) & (8) & (9) \\
\hline
78393 & 162.83676 & 57.32841 & 4.9671 & 46278 & -22.79 & FIR & \lya &  \\
206388 & 163.15096 & 57.26719 & 1.7338 & 13039 & -25.97 & FIR & \civo, 1900\AA &  \\
206427 & 163.04043 & 57.35126 & 3.2798 & 28289 & -24.89 & Xr & \lya, \siivo, \civo & J105209.70+572104.5 \\
206433 & 163.18630 & 57.35626 & 2.8312 & 23704 & -25.60 & Xr & \lya, \siivo, \civo, 1900\AA & J105244.70+572122.4 \\
206445 & 163.51090 & 57.37629 & 1.4058 & 10069 & -22.06 & Xr, FIR & 1900\AA, \mgiio &  \\
206473 & 163.51841 & 57.41341 & 1.9717 & 15268 & -23.70 & FIR & \siivo, \civo, 1900\AA &  \\
206475 & 163.38222 & 57.41506 & 1.9564 & 15123 & -25.57 & Xr & \siivo, \civo, 1900\AA, \mgiio & J105331.73+572454.0 \\
206479 & 163.56356 & 57.41711 & 4.024 & 36098 & -24.00 & Xr & \lya, \civo &  \\
206482 & 163.23815 & 57.41853 & 1.5309 & 11186 & -23.36 & Xr, FIR & 1900\AA, \mgiio & J105257.14+572506.8 \\
206489 & 163.39600 & 57.42837 & 0.7848 & 4901 & -22.13 & Xr & \mgiio, \hb &  \\
206510 & 163.01541 & 57.45189 & 2.4334 & 19735 & -21.60 & FIR, red QSO & \lya, \civo &  \\
206512 & 163.45927 & 57.45264 & 1.7237 & 12946 & -24.88 & Xr & \civo, 1900\AA, \mgiio & J105350.22+572709.5 \\
206531 & 163.28877 & 57.47232 & 1.5698 & 11538 & -24.27 & Xr, FIR & \civo, 1900\AA, \mgiio & J105309.29+572820.4 \\
206557 & 163.10252 & 57.50262 & 1.009 & 6681 & -20.76 & Xr & \mgiio &  \\
206562 & 163.24654 & 57.50832 & 1.676 & 12507 & -22.95 & Xr & \civo, 1900\AA &  \\
206570 & 162.90556 & 57.51196 & 0.8922 & 5739 & -20.46 & Xr & \mgiio &  \\
206579 & 163.41561 & 57.51794 & 0.5855 & 3427 & -22.72 & Xr & \mgiio, \hb, \hg &  \\
206593 & 163.05521 & 57.53933 & 1.873 & 14337 & -24.30 & Xr & \civo, 1900\AA & J105213.25+573221.5 \\
206597 & 163.48816 & 57.54483 & 1.28 & 8967 & -23.25 & Xr, FIR & 1900\AA, \mgiio & J105357.13+573241.4 \\
206623 & 163.27522 & 57.57350 & 2.9514 & 24922 & -26.14 & Xr & \lya, \siivo, \civo, 1900\AA & J105306.04+573424.6 \\
206625 & 162.97661 & 57.57711 & 0.877 & 5619 & -22.30 & Xr, FIR & \mgiio &  \\
206641 & 163.31984 & 57.59742 & 1.2023 & 8298 & -24.83 & Xr & 1900\AA, \mgiio & J105316.76+573550.7 \\
206653 & 163.51091 & 57.61362 & 2.6337 & 21721 & -25.39 & Xr & \lya, \siivo, \civo, 1900\AA &  \\
206666 & 162.90501 & 57.63184 & 1.9387 & 14956 & -24.90 & Xr, FIR, SMG & \siivo, \civo, 1900\AA, \mgiio & J105137.18+573754.7 \\
206667 & 163.26008 & 57.63248 & 1.8837 & 14437 & -24.38 & Xr & \siivo, \civo, 1900\AA, \mgiio & J105302.41+573756.8 \\
206672 & 163.03078 & 57.64392 & 2.7267 & 22652 & -24.30 & Xr & \lya, \siivo, \civo &  \\
206679 & 163.12537 & 57.65379 & 1.437 & 10345 & -23.57 & Xr & 1900\AA\ &  \\
206692 & 163.09724 & 57.68931 & 1.5326 & 11201 & -23.29 & Xr & 1900\AA, \mgiio &  \\
206695 & 163.02548 & 57.69012 & 0.4619 & 2577 & -20.94 & Xr & \mgiio, \hb &  \\
206764 & 162.93636 & 57.46886 & 3.4 & 29534 & -24.10 & Xr & \lya, \siivo, \civo &  \\
\hline
\end{tabular}
\caption{Type 1 AGN sample. 
Columns are as follows: (1) object identifier in the catalog by \citetalias{Gonzalez-Otero2023}, (2) and (3) RA, DEC  (epoch J2000), (4)  spectroscopic $z_{\rm spec}$ reported by \citetalias{Gonzalez-Otero2023}, (5) luminosity distance in Mpc, (6) absolute magnitude in B band computed from the SDSS photometric system following 
\citet{Jordi2006}, (7) object identification reported by \citetalias{Gonzalez-Otero2023}, (8) regions considered for the spectral component fitting, (9) spectra found in the SDSS DR17.  
}
    \label{tab:sample}
\end{table*}

\section{spectral analysis}
\label{sec:fit}


Type 1 AGN provide information on the kinematics of the gas closest to the SMBH. In addition, the diversity of their emission lines, broad permitted and semi-forbidden, and high-, intermediate-, and low-ionization lines allows us to infer the composition and physical properties of the emitting gas. Therefore, careful spectral analysis of each emission line is a fundamental step in optical-UV spectroscopic studies \citep[e.g.][]{sulentic00a, marziani10, Panda2024}. This section defines the main spectral components in the broad and narrow optical and UV spectra to be considered, depending on different kinematic and physical conditions. We also defined the spectral regions selected for our analysis and their spectral components. 

\subsection{Methodological considerations}
\label{sec:meth_considerations}

Spectral analysis of the optical and UV emission lines are performed by separating the virialized broad components (BC) from the wind regions, the narrow lines located in an outer region, the underlying continuum, and the \feii\ pseudo-continuum emission. Our analysis is based on previous works on spectral analysis 
in the optical \citep[e.g.][]{zamfir10, negrete18, Benitez2023, Mengistue2023, Ibarra-Medel2025} and in the UV ranges \citep[e.g.][]{marziani13, negrete14, martinez-aldamaetal18, Garnica2022, Buendia-Rios2023}, which takes into account the different ionization stages of the broad emitting lines. 

\textit{Broad lines}. BELs are considered to emerge from the virialized BLR. We assume that these emission lines have a symmetric profile characteristic of Doppler broadening. However, we also consider previous studies suggesting that the BLR emission line profiles may be Gaussian or Lorentzian \citep[e.g.][see also \citealt{Kollatschny2013} for analysis of line profile broadening simulations]{marziani10}. The choice of the type of profile to use will be constrained by other spectral features, particularly the FWHM, the intensity of the \feii\ in the optical range, and the \civ\ asymmetry in the UV range. BELs with FWHM $ \,<\,$ 4000 \kms, are better fitted with a Lorentzian profile since it describes the central emission and the extended wings characteristic of this profile \citep[see Figure 2 of][and Figure \ref{fig:fits_GTC}]{marziani10}. Usually, these Lorentzian line profiles are accompanied by strong to moderate \feii\ emission in optical and UV emission by prominent \civ\ blue asymmetries. For broader lines with FWHM $ \,>\,$ 4000 \kms, a Gaussian profile adequately represents the virialized component of the BLR  \citep[see Figure 2 in][and Figure \ref{fig:fits_GTC}]{marziani10}. Gaussian profiles are the choice to fit objects with a faint or null \feii\ optical emission and a more symmetric \civ\ profile. We assume that this broad line is the ``core'' component of the BLR and should be centered in its respective restframe with a small range of possible shifts within the range of the spectral resolution.
   
\textit{Blue and redshifted line profiles}. For the case of intermediate-to-high-ionization lines (IILs and HILs), a blueshifted component associated with outflow winds is expected \citep{sulentic07,sulentic17}. This blue component is prominent in the broad lines with FWHM $ \,<\,$ 4000 \kms, specifically in the high ionization lines (HILs) \civ, \heii, and \siiv\ in the UV spectra, and in the narrow lines of \oiii\ in the optical spectra. 
The line profile used to fit for the blue component is an asymmetrical Gaussian profile with FWHM $\,>\,$ 6000 \kms \citep{Proga2000, panda19, Marziani2022}. For broad lines with FWHM $ \,>$ 4000 \kms, a very broad redshifted component (VBC) can also be present in all BELs. The origin of this VBC remains unclear \citep{marziani10}. Some studies suggest that it can be due to optically thin gas present in quasars with low accretion rates that allows matter to approach the inner edge of the accretion disc, creating a region with gas at very high velocities. 
This redshifted, VBC emission with FWHM $\,\sim\,$10,000\,\kms, can become a very prominent component in some spectra that cannot be attributed to other emitting components \citep[e.g.][]{Benitez2022, Buendia-Rios2023}.
    
\textit{Iron emission}. \feii\ emission is strong in the optical range, as seen in the composite quasar spectrum of \cite{vanden01}. This \feii\ emission generates a pseudo-continuum, extending from 4000 to 5400 \AA, and is the product of the multiplet transitions of the ion \citep[e.g.][]{Kovacevic2010,Mengistue2023}. We will consider the optical \feii\ template used in \cite{negrete18} obtained from a high-resolution spectrum of I Zw 1 plus a model of the emission computed by a photoionization code in the \hb\ range \citep[see also][]{marziani09}. The \feii\ emission is also blended and strong in the vicinity of \mgii\ \citep{Sameshima2009,marziani13,Prince2023}. In the UV range around 2800 \AA, we consider \cite{bruhweiler08} \feii\ emission templates computed from {\sc cloudy} simulations \citep{ferlandetal09}. In the range around the 1900 \AA\ blend, \feii\ is not as intense, and we only consider the emission of \feiiuv\ (FeIIUV191) as an isolated component. In contrast, the \feiii\ emission is intense between 1840-2120 \AA\ \citep{negrete12}. For the \feiii\ multiplets, we used the template of \cite{vestergaard01}. 

{\it Narrow lines}. All permitted and 
forbidden narrow emission lines were assumed to be created in the external low-density (log \nh\ $\sim$ 4) narrow line region (NLR). The typical FWHM of these lines is below 1000 \kms, except for a second blueshifted component of \oiii, which can reach up to 2000 \kms\ \citep[the semi-broad SB component][]{negrete18}. We assume that all narrow components (NCs) are emitted in the same region and, therefore, share the same kinematics. So, for each fitting range, the same FWHM and offsets were used for all narrow lines in that region.
    
\textit{Continuum}. Whenever possible, we fitted a single power-law to account for the underlying continuum of the whole spectral range. We use continuum windows around 1280, 1350, 1450, 1700, 2150, 3000, 4440, and 5100 \AA\ \citep[see e.g.][]{francis91}. 
A special treatment for the continuum was considered in the \lya\ range owing to the \lya\ forest \citep{Rauch1998Lyaforest}. A series of absorption lines on the blue side of \lya\ are expected because of the hydrogen clouds located in our line of sight. To set the continuum under the \lya\ range, we assume the continuity of the power-law considering the continuum windows at 1350 \AA\ and 1700 \AA.

\subsection{Spectral fitting}
\label{sec:spec_fitting}

We used the {\sc IRAF-specfit} task \citep{kriss94} to perform the spectral fitting. {\sc specfit} 
simultaneously fits the emission and absorption line components and the continuum models. It requires selecting a set of functional forms to model the emission and absorption line components, the underlying continuum, and the power-law index. Subsequently, {\sc specfit} fits the strong BELs, followed by the narrow components and  \feii\ optical pseudo-continuum. {\sc specfit} can use Gaussian or Lorentzian line profiles by inputting their intensity, central wavelength, and FWHM. Therefore, {\sc specfit} can model complex line systems, such as the complicated continuum of \feii, blended emission lines, and extinction. The fit uses a $\chi^2$ minimization with a Marquardt grid fit algorithm from Numerical Recipes \citep{Press1986, kriss94}.

The considerations set in the previous section allow us to constrain the general parameters of spectral lines. We now describe the spectral features associated with the expected physical conditions for the emission lines, which will help us constrain the number of free parameters in the spectral fits. 
For each spectral range described below, the most intense line was first fitted, if more than one emission line was observed within the spectral range. For objects with $z$ $>$ 2.4, the most intense line is \lya, however, this is also the line with the largest absorption. Thus, for objects with $z$ $>$ 1.28, the dominant line is \civ. 

\textit{\hb}. The spectral range considered was 4440 - 5400 \AA. The continuum was fitted considering the windows at 4440 and 5100 \AA, followed by the \feii\ pseudo-continuum. 
We then model the \hb\ BC, the narrow and SB components of \oiiill\ considering the theoretical 1:3 ratio \citep{Osterbrock1981}, and the NC of \hb. Finally, the fit of \heiiopt\ was included by considering the residuals. For object 206579, the high S/N and spectral coverage allowed the modeling of \hg\ to be done together with the fit of \hb. 

\textit{\mgii.} The fitted range is from 2600 to 3050 \AA. The continuum was fitted considering both the window at 3000 \AA\ and \feii $_{UV}$ emission. 
\mgii\ is a doublet where we consider a line ratio of I(Mg{\sc ii}$\lambda$2796.35)/I(Mg{\sc ii}$\lambda$2803.53) = 0.8 \citep{marziani13,Marziani13b}. Using the \cite{bruhweiler08} \feii\ template gives us a residual around 2950 \AA\ associated with an emission of Fe{\sc i}, which we fitted with a Gaussian component. 
Looking at the residuals, a single \mgii\ NC was considered to achieve the best fit for some objects.
    
\textit{1900\AA\ blend}. The main emissions in the spectral range between 1750 and 2000 \AA\ are three IILs: \ciii, \siiii, \aliii. 
Of particular importance are the \ciii\ and \siiii\ lines, as they are broad semi-forbidden lines and have served to constrain the physical parameters in the emitting region, mainly the density and the ionization parameter \citep{negrete12, negrete13, negrete14, Garnica2022}. The \aliii\ doublet was fitted using two separate components at 1854.6 and 1862.2 \AA\ and a line ratio 1:1.25. In this region, \feiii\ emission is also observed, which is particularly intense in objects where \aliii\ is also intense (see Sections \ref{sec:meth_considerations} and \ref{sec:lineComps_AGNParams}). We also expect the contribution of multiplets of \feii\ as seen in \cite{negrete14}; however, we fitted only the strongest line \feiiuv\ as the rest of the lines of the multiplet are much fainter and buried under the blend. Finally, we consider weaker emissions around the blend such as \siii\ and \niii.
    
\textit{\civ, \heii, and \siiv}. As these three lines share high ionization potentials, they were fitted with the same BC profile, FWHM, shift, and several complementary components that share the same profile (see Sec. \ref{sec:fit}). \civ\ and \heii\ were fitted at the same time because the blue side of \heii\ is usually blended with the red wing of \civ. We model \siiv\ separately because in high-z objects, it lies on the blue edge of the spectra and becomes noisy or with large flux uncertainties. 
It is well known that the \siiv\ doublet is blended with O${\rm IV}\lambda$]1402. However, the O${\rm IV}\lambda$]1402 BC emission is expected to be negligible in comparison to the \siivo\ emission \citep[][]{Juarez2009, negrete12, negrete14}. 
We fitted the \civ\ line in the range 1450-1700 \AA, setting the BC of \civ\ in its restframe, to reach its maximum intensity. Based on the residuals, a blue and/or a rdshift component as described in Sec.~\ref {sec:meth_considerations} I is considered. Then we fitted \heii\ and other faint lines such as \oiuv\ and \niv.
SiIV: The $\lambda$1400 blend was fitted similarly to the \civ, in the range of 1350-1450 \AA, using those found in \civ as initial conditions. We used a BC considering the doublet with individual lines at 1402 and 1394 \AA\ and a flux ratio $F$(1394) = 0.91 $F$(1402). In agreement with the \civ\ fit, a blue and/or redshifted component was fitted for some objects.
    
\textit{\lya\ region.} In addition to the \lya\ components that were fitted based on the emission components and line profiles of \civ, we also consider modeling \nv, \siiiuv, \oiuv, and \ciiuv\ lines to obtain a more robust fitting. We used Gaussian profiles to model these extra components, fixing the line shifts and widths of the faintest components to have the same values. The fitting range used is between 1180 and 1350 \AA. For objects with absorption produced by the \lya\ forest, the range was shortened to 1210-1350 \AA, and some absorption lines were fitted using a Gaussian model to recover the broad line.
    
\section{AGN Derived Parameters}
\label{sec:lineComps_AGNParams}

\begin{figure*}
  \sbox0{\begin{tabular}{@{}cc@{}}
    \includegraphics[width=0.25\linewidth]{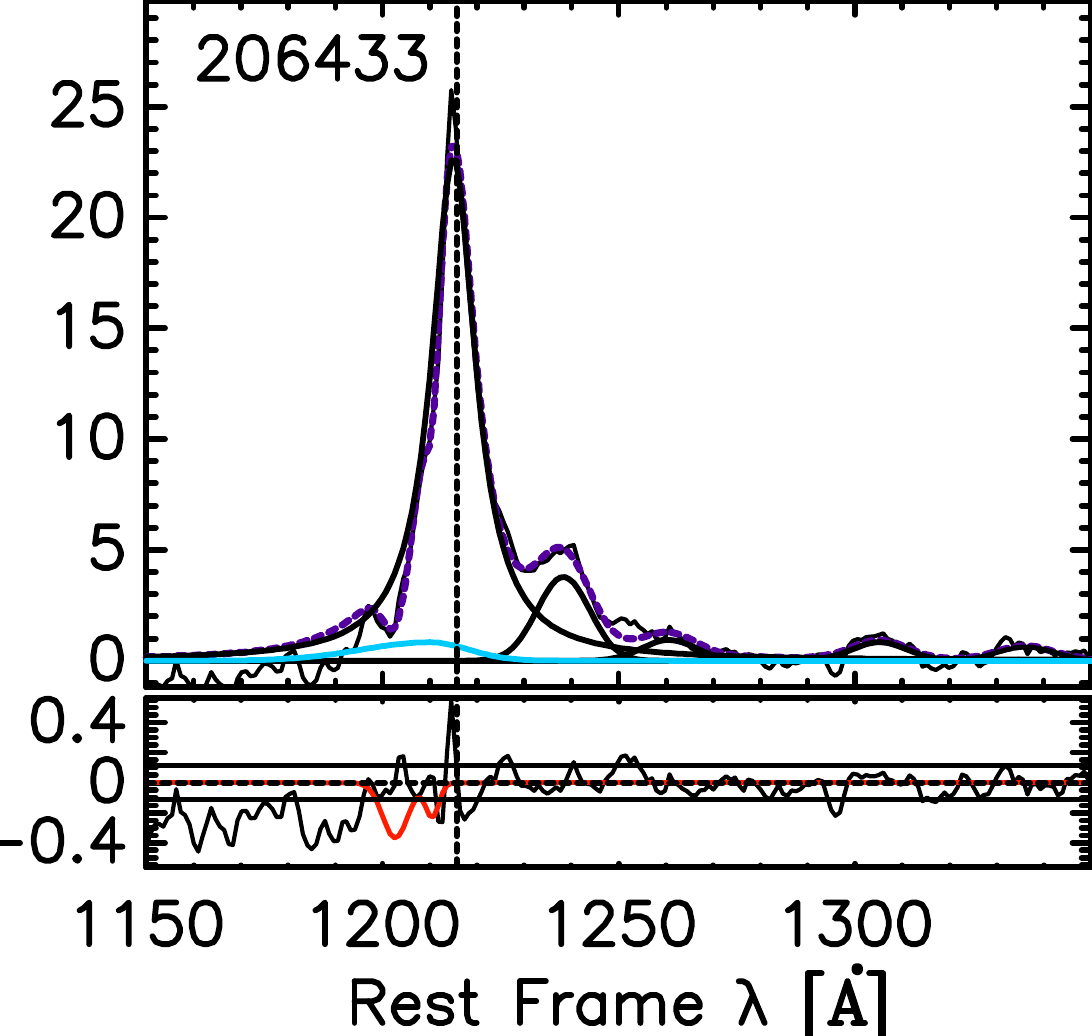}
    \includegraphics[width=0.25\linewidth]{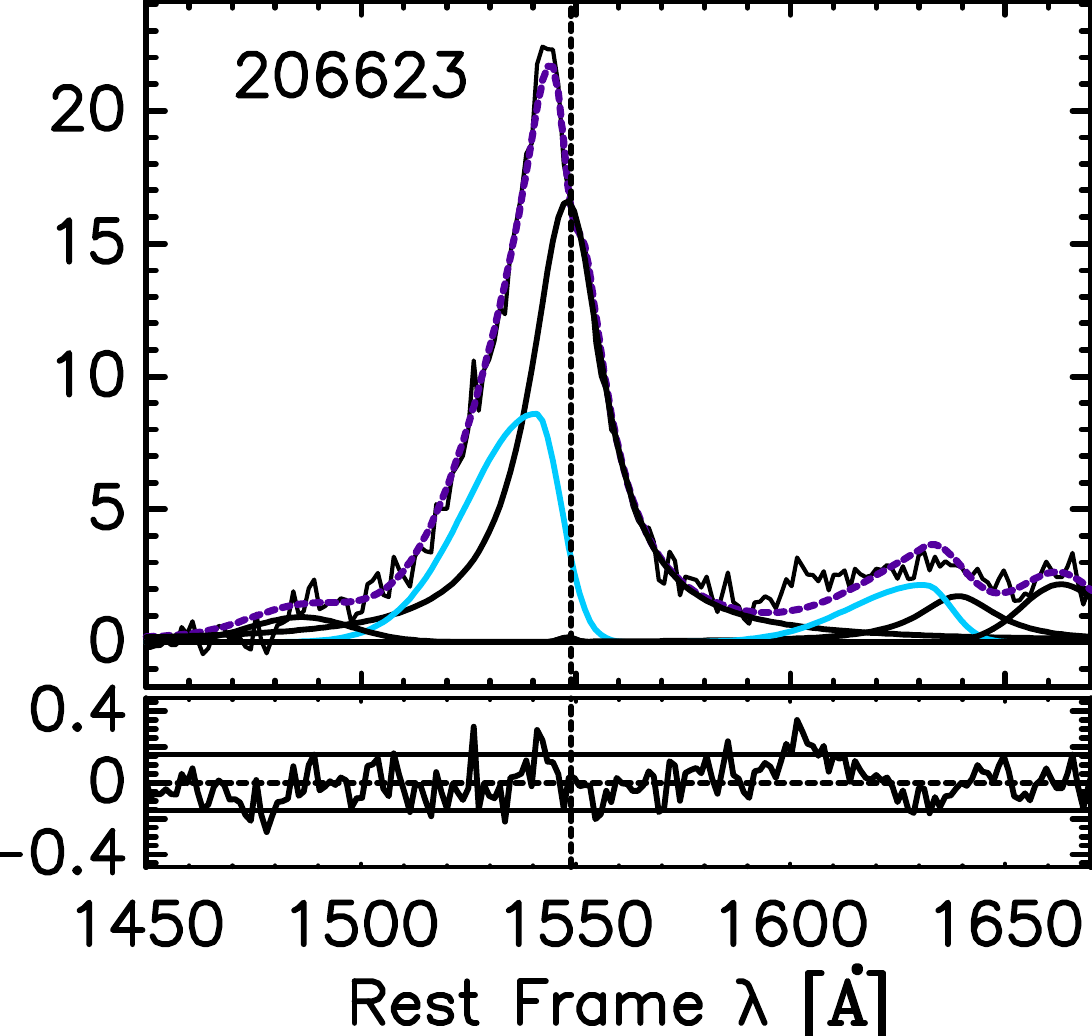}
    \includegraphics[width=0.25\linewidth]{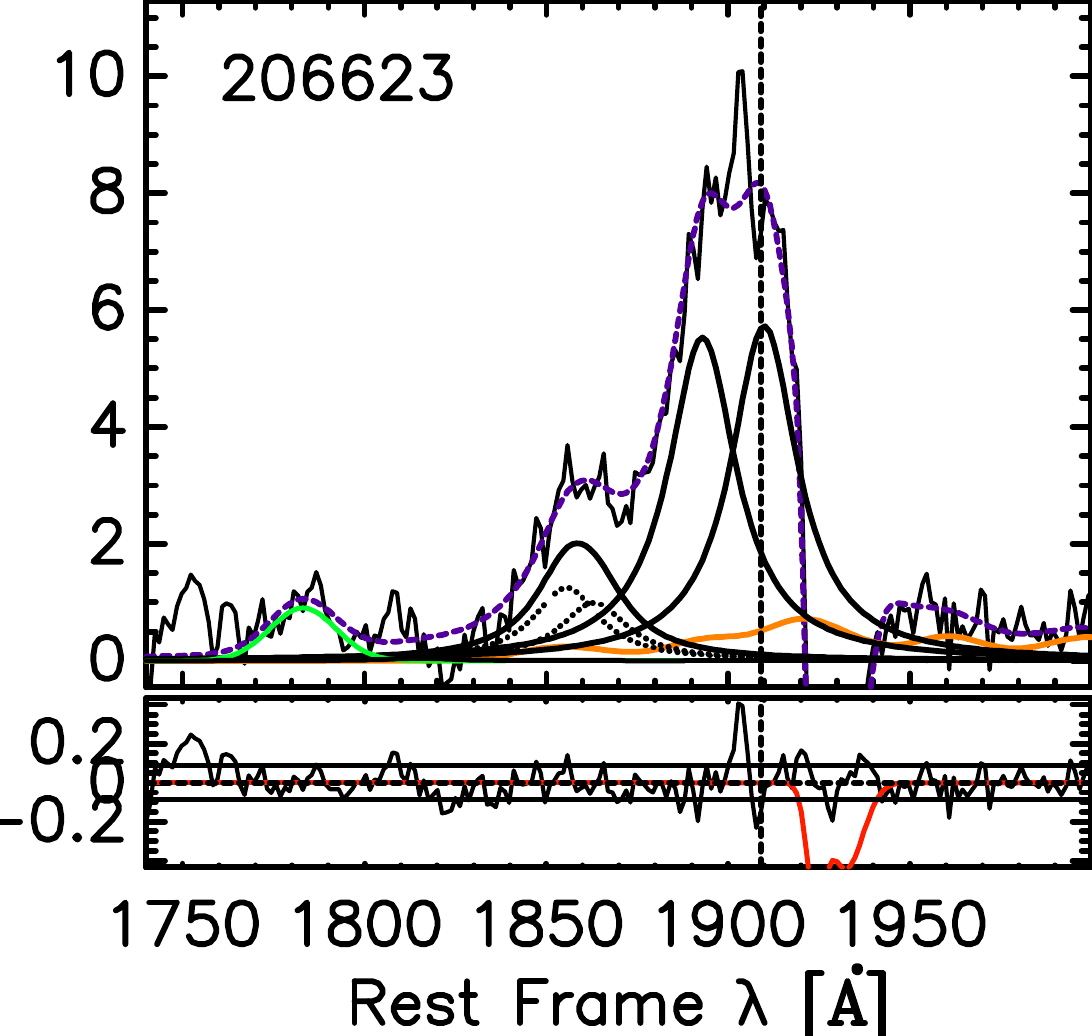} 
    \includegraphics[width=0.25\linewidth]{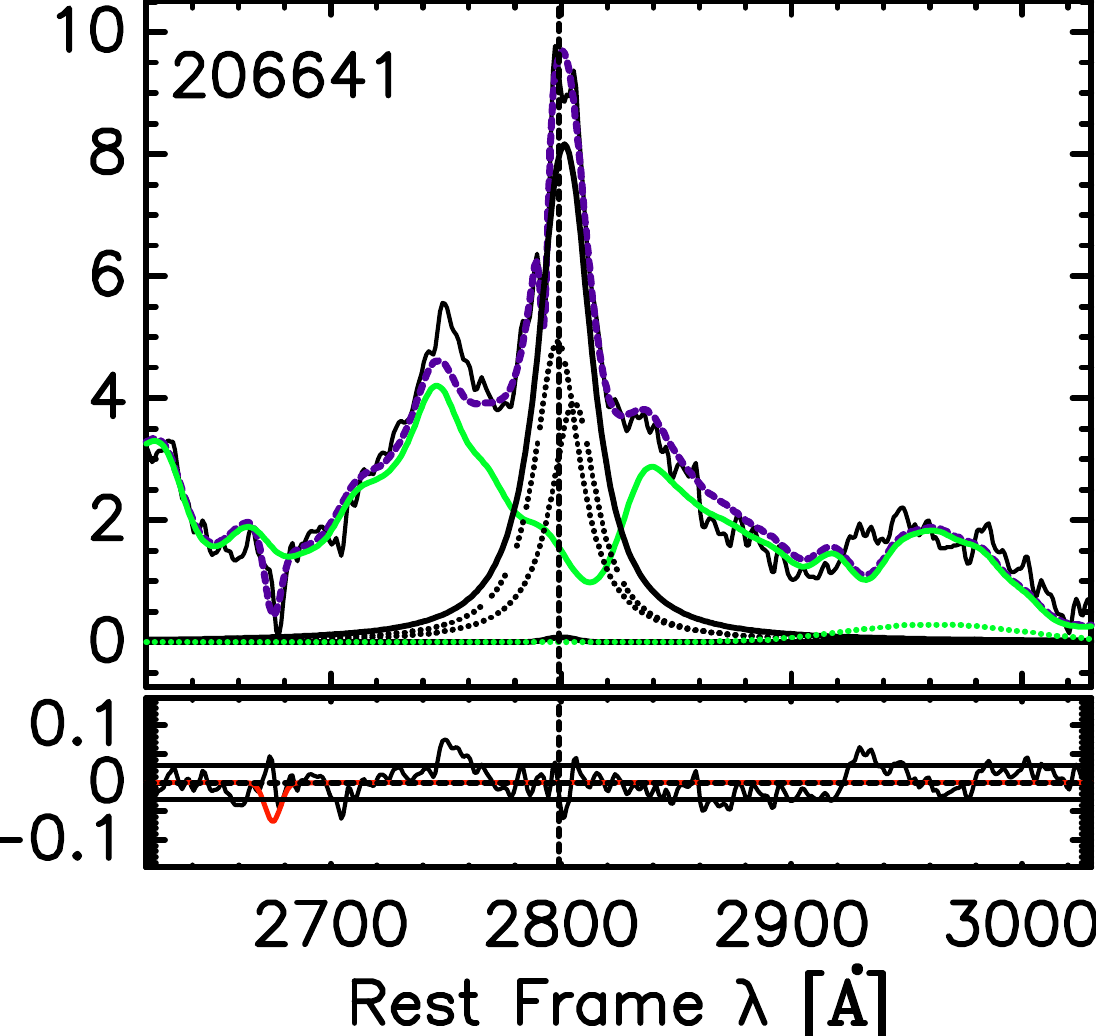}
    \includegraphics[width=0.25\linewidth]{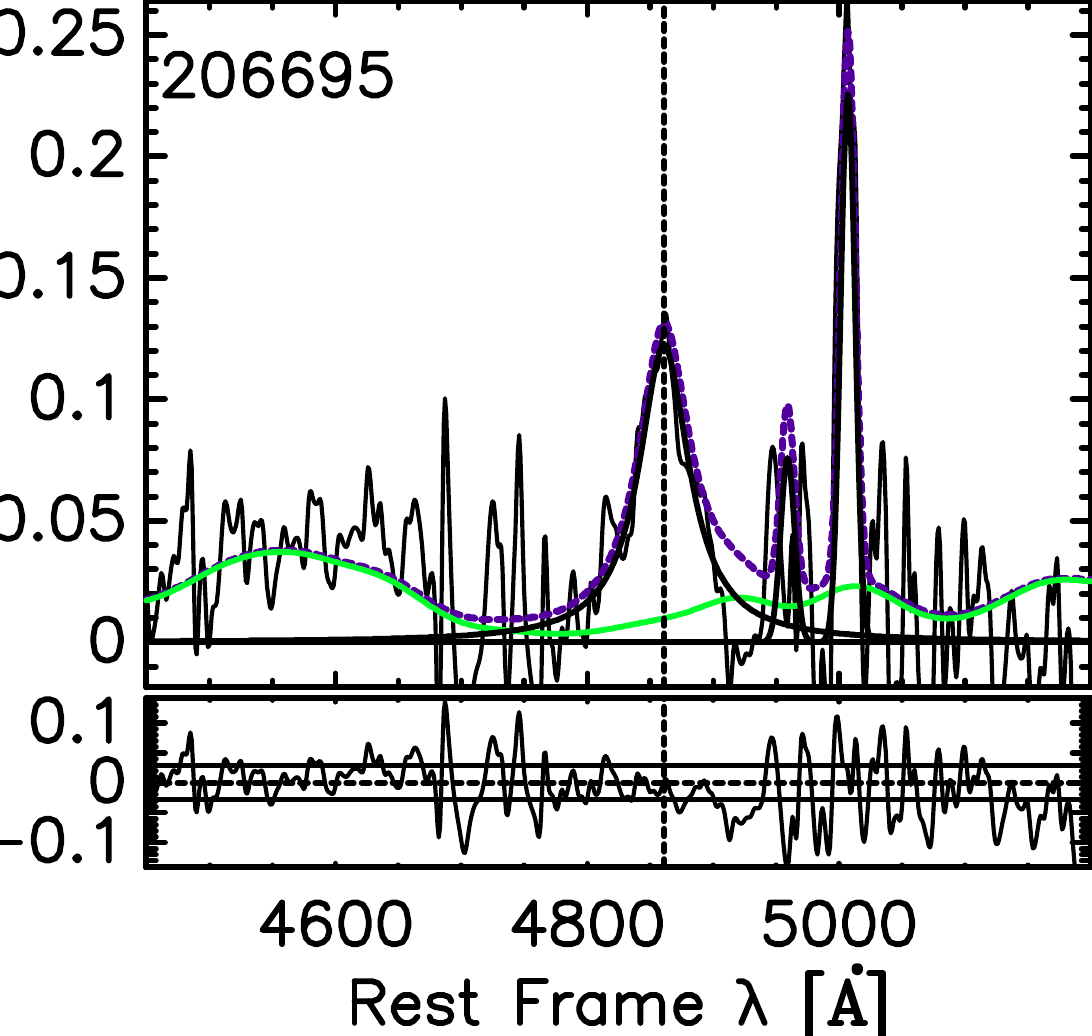}\\
    \includegraphics[width=0.25\linewidth]{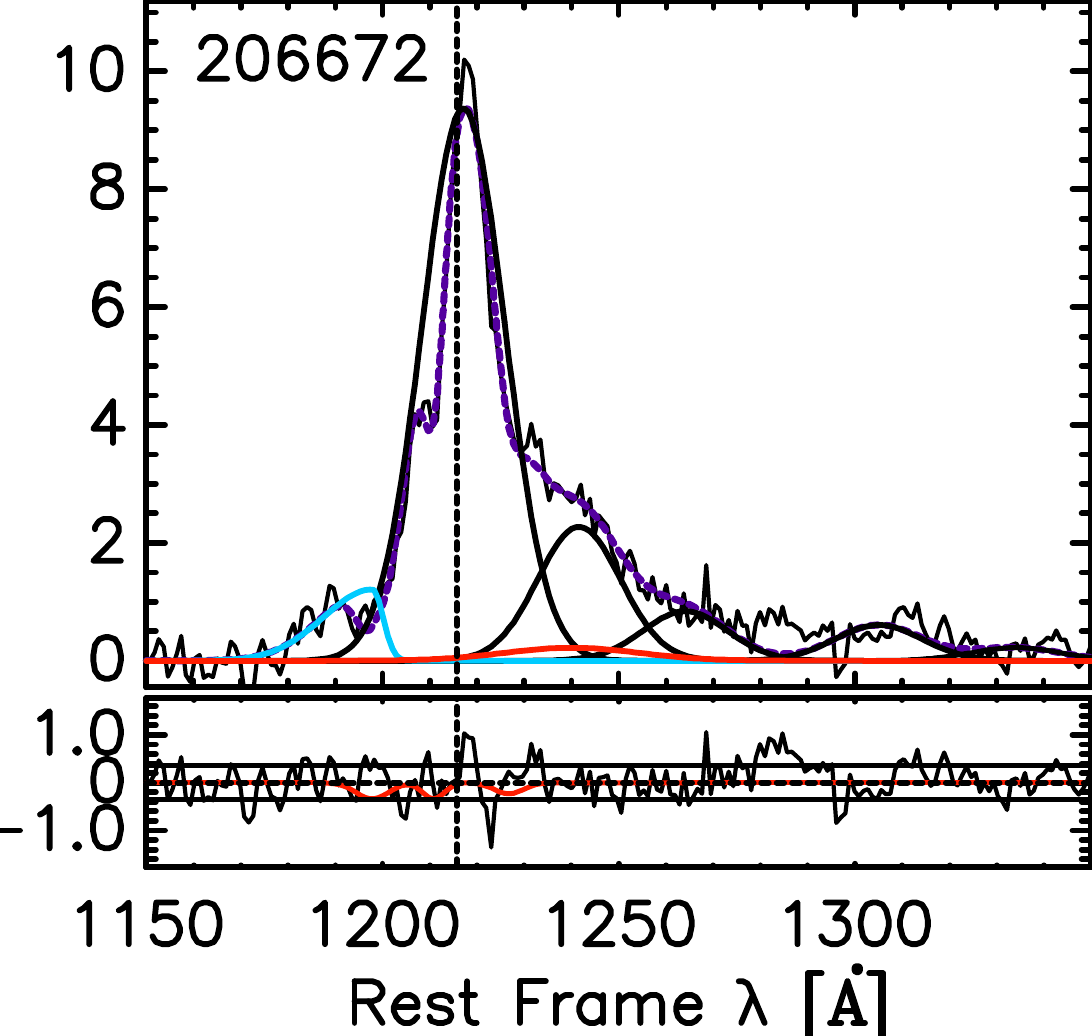}
    \includegraphics[width=0.25\linewidth]{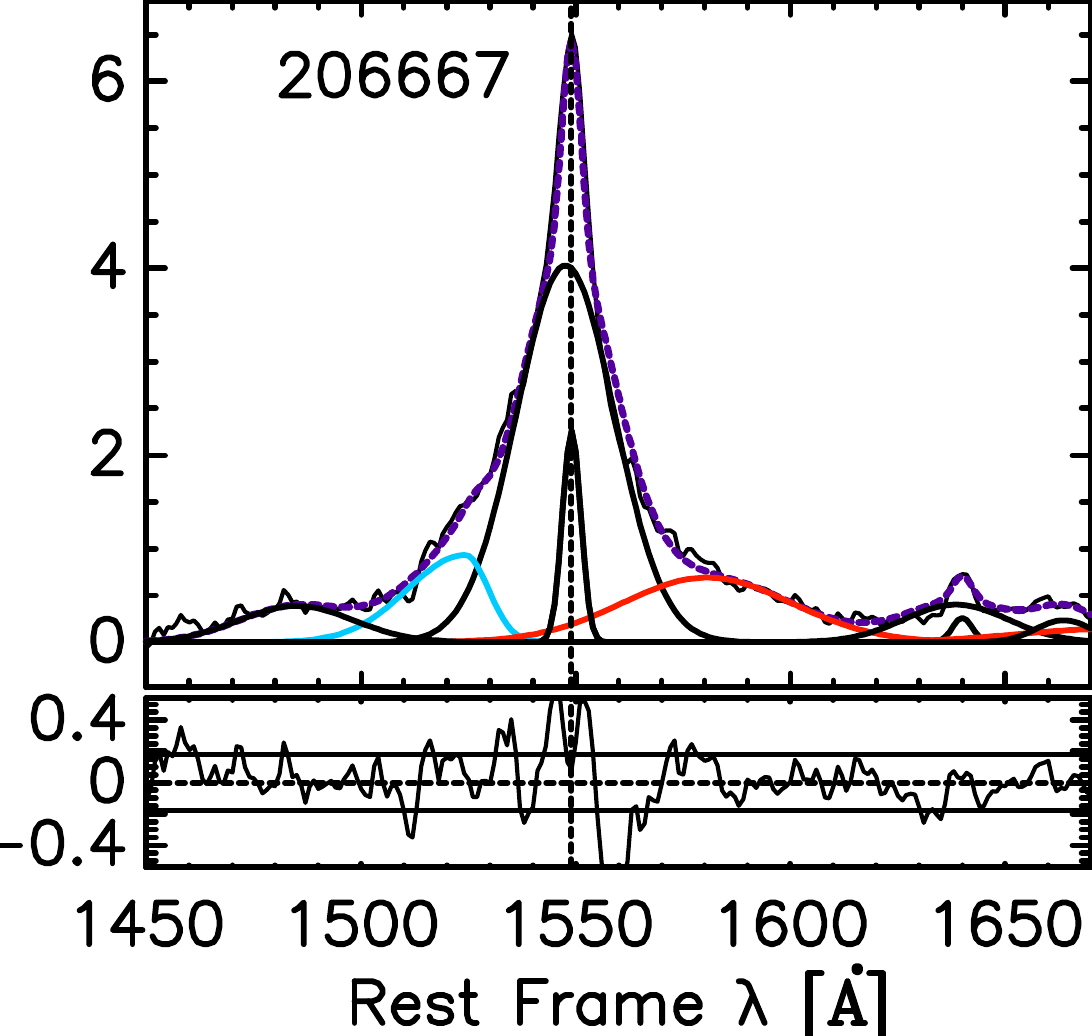}    
    \includegraphics[width=0.25\linewidth]{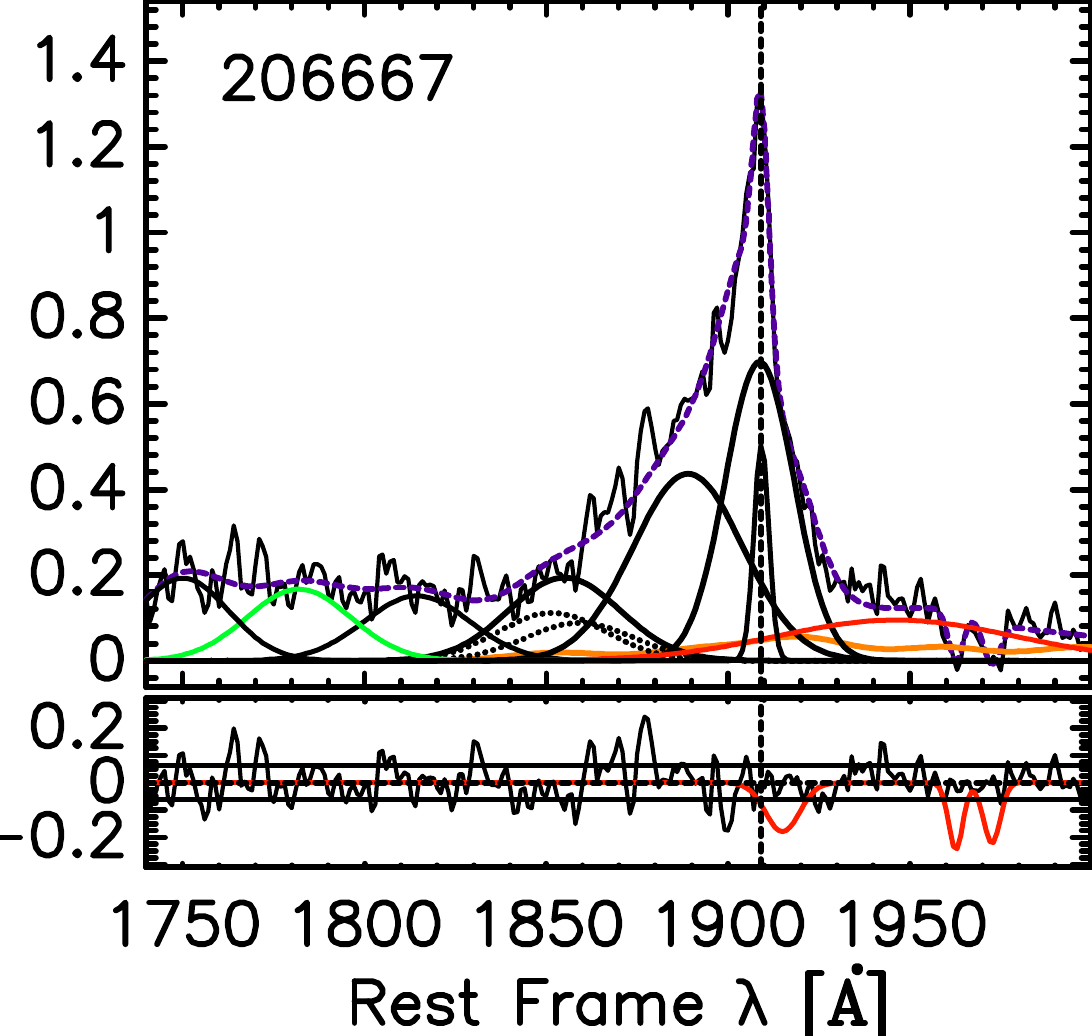}   
    \includegraphics[width=0.25\linewidth]{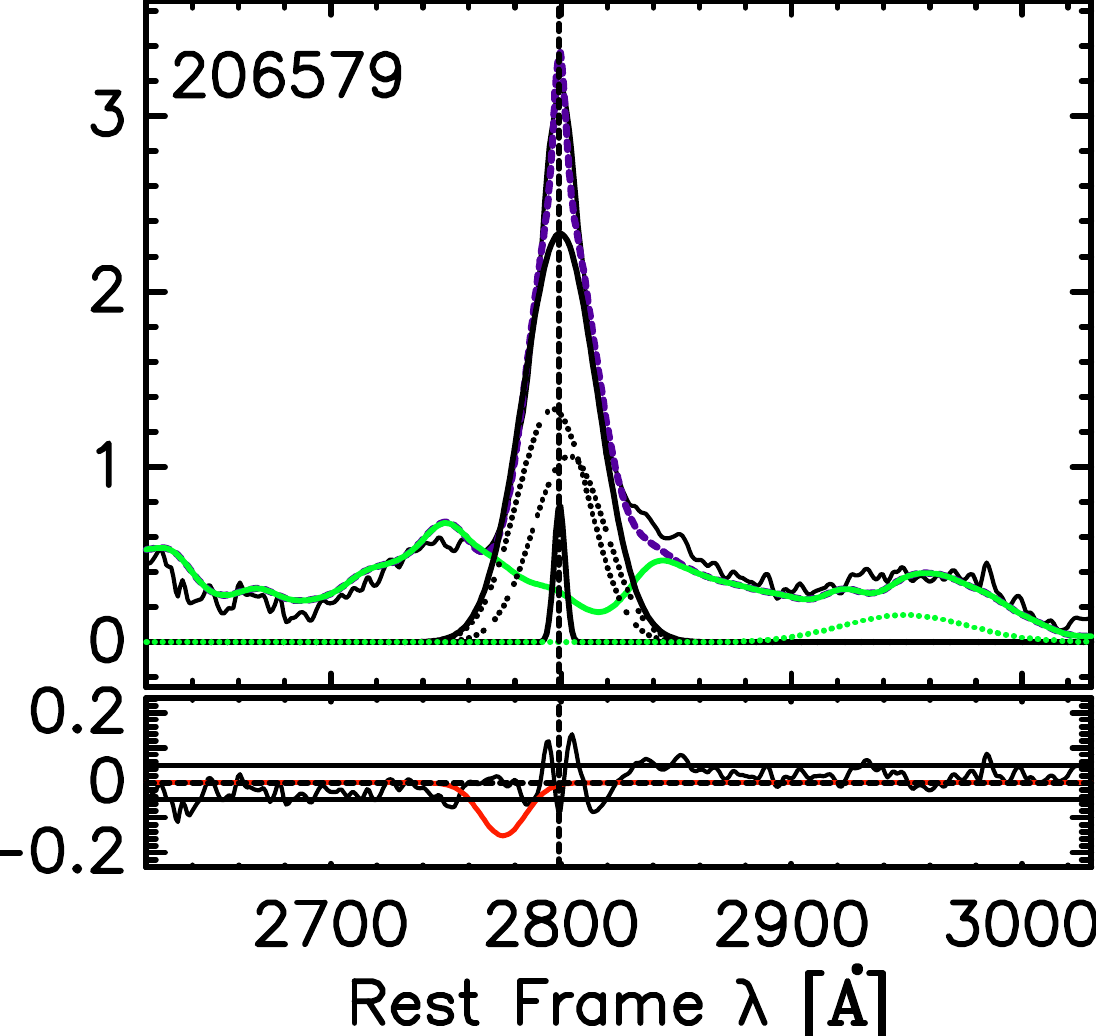}
    \includegraphics[width=0.25\linewidth]{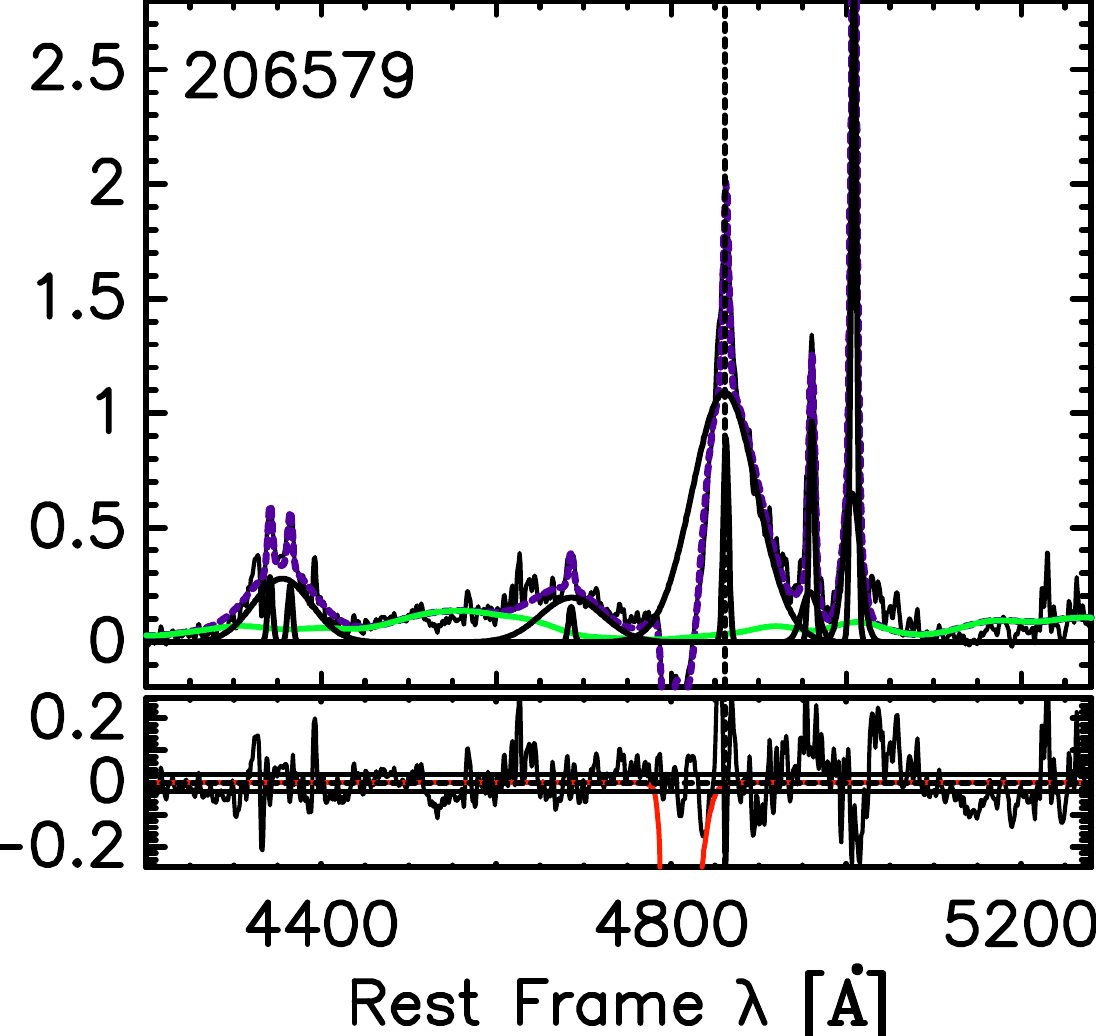}
  \end{tabular}}
    \rotatebox{90}{\begin{minipage}[c][\textwidth][c]{\wd0}
    \usebox0
    \caption{Examples of spectral fitting using Lorentzian (upper panels) and Gaussian (lower panels) profiles for \lya, \civ, 1900\AA\ blend, \mgii, and \hb. The ordinate is the restframe wavelength, while the abscissa is the flux in units 10$^{-16}$ ergs s$^{-1}$ cm$^{-2}$ \AA$^{-1}$, considering the continuum subtracted. The thin black line represents the observed spectra, thick black lines represent the BC, blueshifted components are in blue, VBC emissions are in red, and narrow lines are in gray. The purple dashed line is the sum of all the fitted components. The vertical dashed line is the restframe. 
    Green continua in \mgii\ and \hb\ are the FeII template. In the 1900\AA\ blend, the FeIII template is shown in orange. The lower panels are the residuals, with red lines showing the absorption lines considered in the fit.
    \label{fig:fits_GTC}
} 
  \end{minipage}}
\end{figure*}

In this section, we report the line profile parameters of our sample along with the derived parameters obtained from the spectral fitting analysis. Individual parameters like $\lambda_{cent}$, $f_{line}$, FWHM, velocity shifts, EW, measured in the spectral regions described in Sec. \ref{sec:spec_fitting}, as well as the individual computations of the derived parameters (\lbol, \mbh, and \redd\footnote{{\redd = \lbol/\ledd} is the Eddington ratio, the ratio of the bolometric luminosity (\lbol) and the Eddington luminosity (\ledd\ = 1.26 $\times 10^{8}$ M/M$_\odot$ \ergs).}) and the figures showing the fit done in each spectral region, are available by contacting the corresponding author upon request.
Figure ~\ref{fig:fits_GTC} shows examples of each spectral range covering \lya, \siiv, \civ, 1900 \AA\ blend, \mgii, and \hb\ for Lorentzian and Gaussian broad component profiles. 

The SMBH masses (\mbh) were estimated using different calibrators compiled from the literature, using virialized BC in different regions of the optical-UV spectrum. We compiled the $\alpha$ and $\beta$ coefficients of the equation

\begin{equation}
\label{eq:mbh}
    \small
    {\rm log M_{BH,Line}} =  {\rm log} \left\{ \left[ \frac{({\rm FWHM \left( Line)\right)}}{1000\ {\rm km s}^{-1}}\right]^2 \left[ \frac{\lambda {\rm L_\lambda \left(Cont\right)}}{10^{44}\ {\rm ergs\ s}^{-1}}\right]^\alpha \right\} + \beta
\end{equation}

\noindent from the works of \citet[][VP06]{Vestergaard2006} for \hb, \citet[][TN12]{Trakhtenbrot2012} for \mgii, \citet[][BR23]{Buendia-Rios2023} for \aliii, and \citet[][M19]{Marziani2019} for \civ. 
It is important to note that for the \mbh\ computations considering \civ, M19 isolate the virialized BC from the wind blueshifted component.
Equation \ref{eq:mbh} computes the \mbh\ using different virial estimators, where $\lambda L_\lambda$ is the continuum luminosity Cont(\AA), in a specific continuum window, and FWHM(line) is from the reference line. Table \ref{tab:MBHcoeff} resumes the values used in eq. \ref{eq:mbh}. The references for the bolometric correction (B.C.) are \citet[][S17]{sulentic17}, \citet[][MS14]{MS14}, and \citet[][R06]{richards06}. 
\citet{Krawczyk2013} propose later a BC = 2.75 in the UV at 2500\AA, using an integrated spectral energy distribution from 1 $\mu$m to 2 keV. However, the difference with R06 is not large.

\begin{table}
    \centering
    \begin{tabular}{lcccccc}
    \hline
    Line & Cont(\AA) & B.C. & Ref(B.C.) & $\alpha$ & $\beta$ & Ref(\mbh) \\
    (1) & (2) & (3) & (4) & (5) & (6) & (7) \\
    \hline
    \hb  & 5100 & 10.33 & R06 & 0.50 & 6.91 & VP06 \\
    Mg{\sc ii} & 3000 & 5.15 & R06 & 0.62 & 6.75 & TN12\\
    Al{\sc iii} & 1700 & 6.3 & MS14 & 0.58 & 3.24 & BR23 \\
    C{\sc iv} & 1350 & 3.5 & S17 & 0.63 & 3.35 & M19 \\
    \hline
    \end{tabular}
    \caption{Values 
    used for computing the \mbh\ using eq. \ref{eq:mbh}. 
    }
    \label{tab:MBHcoeff}
\end{table}

We also computed the Eddington ratio \redd\ values, a fundamental parameter underlying the spectral differences in AGN \citep[Sec. \ref{sec:LHinQMS}, see also e.g.,][]{marziani01}. 
Table \ref{tab:derived_parameters} reports the continuum flux values and the derived properties of the sample. 
For FWHM(BC), log\lbol, log\mbh, and \ledd, the reported errors include the standard deviation of individual computations. 
The ranges and median values $\mu$ of log\lbol\ (in \ergs) are: 
\begin{itemize}
    \item log\lbol$_{,1350}$ = 46.11--47.87, $\mu$ = 47.23
    \item log\lbol$_{,1700}$ = 45.98--47.95, $\mu$ = 46.84
    \item log\lbol$_{,3000}$ = 44.67--47.30, $\mu$ = 45.54
    \item log\lbol$_{,5100}$ = 45.13--45.80, $\mu$ = 45.78
\end{itemize}
In this log\lbol\ distribution, we see a trend in the increasing log\lbol\ median values towards the smallest wavelengths, except for log\lbol(5100\AA). 
This is because the 5100\AA\ range was obtained from three objects, whereas the 3000\AA\ range from 15 objects spanning a wider \lbol\ range.


\begin{table*}
    \scriptsize
    \centering
    \begin{tabular}{lcccccccccccc}
\hline
ObjID & S/N & Profile & FWHM(BC) & $f$(1450) & $f$(1700) & $f$(3000) & $f$(5100) & log\lbol & log\mbh & log\redd \\
(1) & (2) & (3) & (4) & (5) & (6) & (7) & (8) & (9) & (10) & (11) \\
\hline
78393 & 29 & L & 2057 $\pm$ 261 & 7.64 $\pm$ 1.40 &    &    &    & 47.87 $\pm$ 0.16 & 9.31 $\pm$ 0.31  & 0.56 $\pm$ 0.02 \\
206388 & 12 & L & 2833 $\pm$ 174 & 1.63 $\pm$ 0.20 & 1.36 $\pm$ 0.13 &    &    & 46.35 $\pm$ 0.15 & 8.41 $\pm$ 0.17  & -0.07 $\pm$ 0.05 \\
206427 & 18 & L & 3283 $\pm$ 321 & 3.10 $\pm$ 0.46 & 2.84 $\pm$ 0.43 &    &    & 47.32 $\pm$ 0.20 & 9.50 $\pm$ 0.22  & -0.39 $\pm$ 0.16 \\
206433 & 25 & L & 3564 $\pm$ 240 & 5.02 $\pm$ 0.92 & 4.18 $\pm$ 0.19 &    &    & 47.36 $\pm$ 0.18 & 9.25 $\pm$ 0.16  & 0.08 $\pm$ 0.04 \\
206445 & 37 & G & 6464 $\pm$ 225 &    & 3.15 $\pm$ 0.14 & 1.26 $\pm$ 0.04 &    & 46.49 $\pm$ 0.05 & 9.29 $\pm$ 0.07  & -0.85 $\pm$ 0.09 \\
206473 & 21 & L & 3395 $\pm$ 497 & 1.18 $\pm$ 0.12 & 1.11 $\pm$ 0.07 &    &    & 46.37 $\pm$ 0.12 & 8.56 $\pm$ 0.18  & -0.28 $\pm$ 0.06 \\
206475 & 41 & G & 5635 $\pm$ 399 & 5.37 $\pm$ 0.47 & 5.38 $\pm$ 0.25 & 3.12 $\pm$ 0.11 &    & 47.12 $\pm$ 0.13 & 9.53 $\pm$ 0.12  & -0.42 $\pm$ 0.06 \\
206479 & 10 & L & 2764 $\pm$ 229 & 5.56 $\pm$ 0.92 & 4.63 $\pm$ 0.56 &    &    & 47.76 $\pm$ 0.20 & 9.33 $\pm$ 0.24  & 0.20 $\pm$ 0.05 \\
206482 & 16 & L & 3234 $\pm$ 748 & 2.43 $\pm$ 0.31 & 1.95 $\pm$ 0.13 & 1.33 $\pm$ 0.13 &    & 46.49 $\pm$ 0.12 & 8.68 $\pm$ 0.29  & -0.29 $\pm$ 0.06 \\
206489 & 17 & G & 9546 $\pm$ 376 &    &    & 0.87 $\pm$ 0.01 & 0.42 $\pm$ 0.03 & 45.69 $\pm$ 0.08 & 9.25 $\pm$ 0.07  & -1.70 $\pm$ 1.09 \\
206510 & 6 & L & 2245 $\pm$ 145 & 0.56 $\pm$ 0.10 & 0.34 $\pm$ 0.07 &    &    & 46.18 $\pm$ 0.26 & 8.25 $\pm$ 0.26  & -0.15 $\pm$ 0.11 \\
206512 & 27 & G & 5901 $\pm$ 414 & 1.56 $\pm$ 0.14 & 1.12 $\pm$ 0.07 & 0.74 $\pm$ 0.07 &    & 46.36 $\pm$ 0.19 & 9.13 $\pm$ 0.15  & -0.85 $\pm$ 0.22 \\
206531 & 31 & G & 4750 $\pm$ 304 & 3.33 $\pm$ 0.23 & 3.43 $\pm$ 0.09 & 1.88 $\pm$ 0.30 &    & 46.67 $\pm$ 0.21 & 9.12 $\pm$ 0.15  & -0.55 $\pm$ 0.12 \\
206557 & 17 & G & 4296 $\pm$ 99 &    &    & 0.57 $\pm$ 0.05 &    & 45.67 $\pm$ 0.07 & 8.61 $\pm$ 0.11  & -1.05 $\pm$ 0.39 \\
206562 & 13 & G & 9483 $\pm$ 232 & 1.17 $\pm$ 0.17 & 0.61 $\pm$ 0.05 &    &    & 46.07 $\pm$ 0.16 & 9.32 $\pm$ 0.15  & -1.30 $\pm$ 0.78 \\
206570 & 12 & G & 6098 $\pm$ 362 &    &    & 0.12 $\pm$ 0.01 &    & 44.85 $\pm$ 0.08 & 8.40 $\pm$ 0.15  & -1.70 $\pm$ 1.95 \\
206579 & 37 & G & 4672 $\pm$ 86 &    &    & 2.25 $\pm$ 0.08 & 0.79 $\pm$ 0.02 & 45.73 $\pm$ 0.05 & 8.65 $\pm$ 0.05  & -1.00 $\pm$ 0.13 \\
206593 & 33 & G & 11788 $\pm$ 281 & 2.29 $\pm$ 0.29 & 1.91 $\pm$ 0.06 &    &    & 46.58 $\pm$ 0.12 & 9.80 $\pm$ 0.10  & -1.22 $\pm$ 0.51 \\
206597 & 28 & G & 6621 $\pm$ 160 &    &    & 1.00 $\pm$ 0.03 &    & 46.17 $\pm$ 0.03 & 9.35 $\pm$ 0.05  & -1.30 $\pm$ 0.26 \\
206623 & 47 & L & 3570 $\pm$ 141 & 13.21 $\pm$ 0.43 & 11.32 $\pm$ 0.28 &    &    & 47.83 $\pm$ 0.04 & 9.60 $\pm$ 0.11  & 0.21 $\pm$ 0.01 \\
206625 & 18 & G & 4234 $\pm$ 90 &    &    & 0.60 $\pm$ 0.03 &    & 45.54 $\pm$ 0.04 & 8.52 $\pm$ 0.06  & -1.10 $\pm$ 0.22 \\
206641 & 66 & L & 2466 $\pm$ 99 &    &    & 15.59 $\pm$ 0.32 &    & 47.30 $\pm$ 0.02 & 9.18 $\pm$ 0.06  & 0.01 $\pm$ 0.01 \\
206653 & 19 & G & 5821 $\pm$ 412 & 3.92 $\pm$ 0.25 & 3.64 $\pm$ 0.13 &    &    & 47.20 $\pm$ 0.07 & 9.49 $\pm$ 0.11  & -0.35 $\pm$ 0.04 \\
206666 & 29 & G & 7381 $\pm$ 436 & 3.60  0.16 & 3.12 $\pm$ 0.10 &    &    & 46.82 $\pm$ 0.05 & 9.65 $\pm$ 0.07  & -0.89 $\pm$ 0.10 \\
206667 & 13 & G & 5480 $\pm$ 763 & 0.94 $\pm$ 0.09 & 0.55 $\pm$ 0.06 & 0.67 $\pm$ 0.08 &    & 46.17 $\pm$ 0.24 & 9.02 $\pm$ 0.26  & -0.85 $\pm$ 0.28 \\
206672 & 15 & G & 5260 $\pm$ 583 & 1.75 $\pm$ 0.11 & 1.33 $\pm$ 0.12 &    &    & 46.84 $\pm$ 0.10 & 9.32 $\pm$ 0.18  & -0.66 $\pm$ 0.12 \\
206679 & 22 & G & 5500 $\pm$ 277 &    & 0.70 $\pm$ 0.08 &    &    & 45.98 $\pm$ 0.05 & 8.68 $\pm$ 0.16  & -0.80 $\pm$ 0.16 \\
206692 & 18 & L & 3963 $\pm$ 176 & 2.15 $\pm$ 0.18 & 3.16 $\pm$ 0.10 & 1.03 $\pm$ 0.25 &    & 46.54 $\pm$ 0.23 & 8.90 $\pm$ 0.18  & -0.40 $\pm$ 0.14 \\
206695 & 13 & L & 2435 $\pm$ 379 &    &    & 0.38 $\pm$ 0.01 & 0.33 $\pm$ 0.02 & 44.90 $\pm$ 0.08 & 7.59 $\pm$ 0.18  & -0.80 $\pm$ 0.14 \\
206764 & 15 & L & 2390 $\pm$ 128 & 2.75 $\pm$ 0.79 &    &    &    & 47.16 $\pm$ 0.25 & 9.06 $\pm$ 0.33  & 0.07 $\pm$ 0.11 \\
\hline
\end{tabular}
    \caption{Derived parameters. 
    Col. (1) is the object identification. Col. (2) is the largest S/N value around the given continuum window of the following four columns. 
    Col. (3) is the adopted line profile, 
    Col. (4) are the FWHM average values of the BCs in \kms. Cols. (5-8) is the continuum flux in units 10$^{-16}$ erg$^{-1}$ s$^{-1}$ cm$^{-2}$ \AA$^{-1}$. Cols. (9,10) are the average values of the logarithm of the \lbol\ and \mbh, and Col. (11) is the average value of the \redd.}
    \label{tab:derived_parameters}
\end{table*}

\begin{figure}
    \centering
    \includegraphics[width=0.99\linewidth]{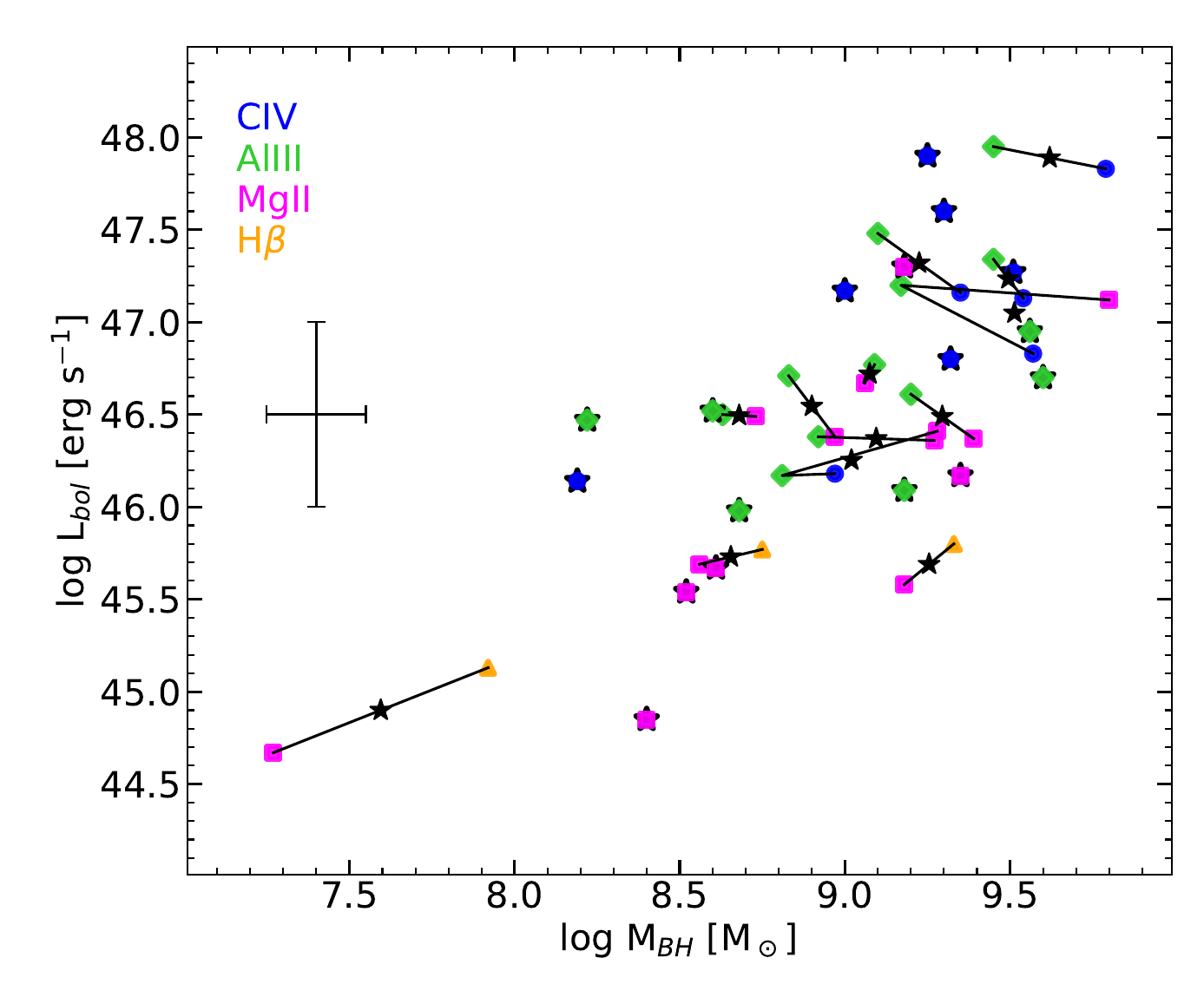}
    \includegraphics[width=0.99\linewidth]{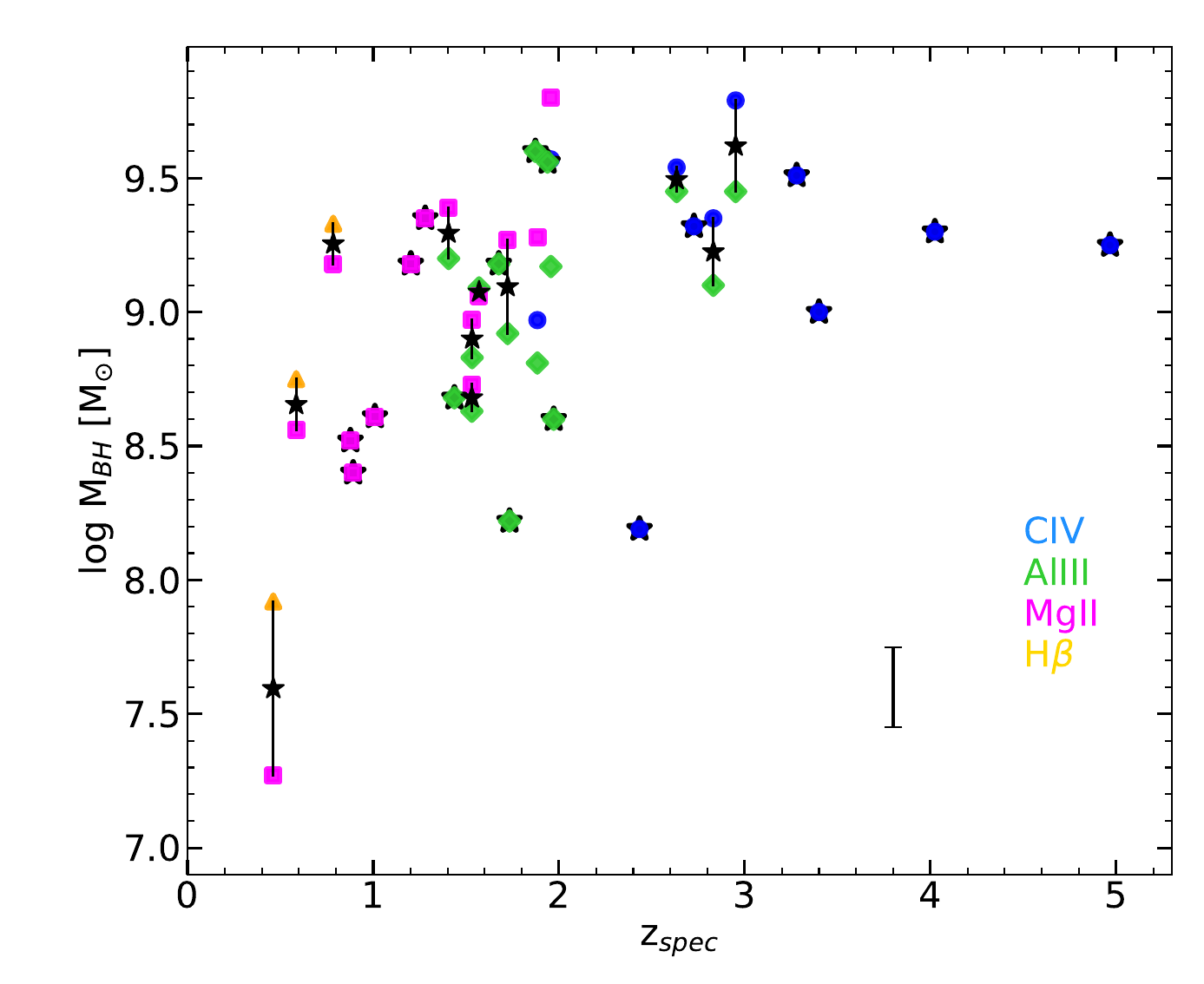}
    \includegraphics[width=0.99\linewidth]{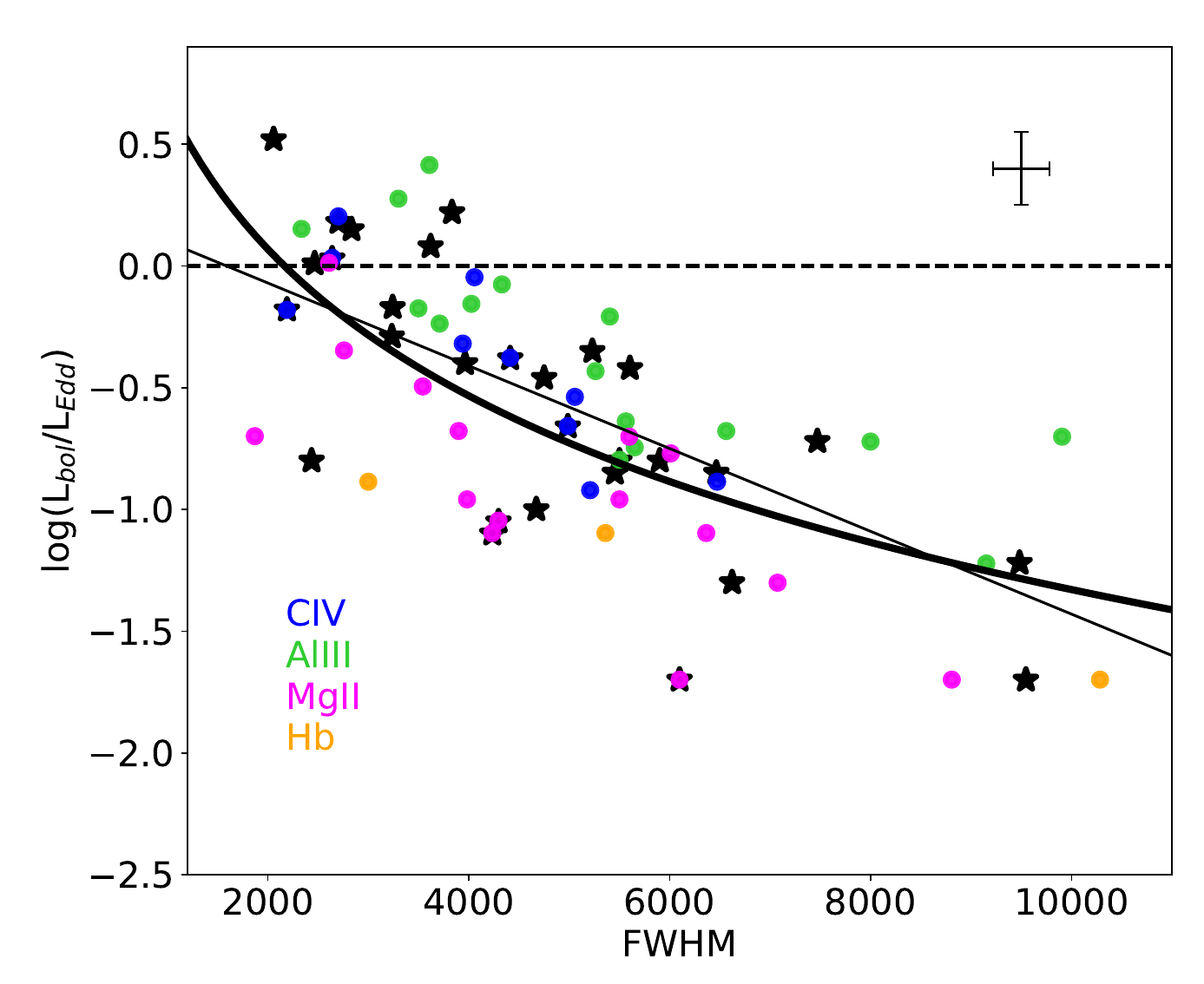}
        \caption{Distributions of log\mbh\ - log\lbol\ (\textit{upper panel}), log\mbh\ in the $z$ range (\textit{middle panel}), and log\redd\ - FWHM(BC) (\textit{lower panel}). In the lower panel, the dashed horizontal line corresponds to \redd\ = 1, the solid thin line represents an anti-correlation between both quantities, and the solid thick line is the expected correlation of \redd\ and the FWHM}. The average error bars are also shown. 
    \label{fig:MBH-z}
\end{figure}

Figure~\ref{fig:MBH-z} presents the dispersion of \mbh\ as a function of log\lbol\ of the individual virial estimators and individual continuum luminosities. 
We also show the distribution of \mbh\ in the redshift range and the relationship between \redd\ and the FWHM of the BCs. 
Colored dots are the individual computations for each virial estimator. We used star symbols to highlight the average values. 
For objects with a single value, the colored dot is superimposed on the star, where the lines connecting the dots to the stars show the values associated with an object. 
We observed a clear trend in the \redd\ vs. FWHM plot and derived the anti-correlation using the FWHM normalized by 1000 \kms: 
\begin{equation}
\label{eq:Redd_FWHM}
    \mathrm{log} R_{\rm Edd} = -0.17 \, \mathrm{FWHM}_{1000}(BC) + 0.27
\end{equation}
with a correlation coefficient r$_p$=-0.7. 
The correlation described above is expected as the \mbh\ that is used to compute \redd\ uses the FWHM values. This expected trend is also shown in the lower panel of Fig. \ref{fig:MBH-z}. Both relations, expected and derived, are consistent with the sample distribution, within the errors. However, previous works using \mbh\ derived from X-ray data show a similar trend, which means. 
that the \redd\ - FWHM is a statistically true correlation \citep{marziani01}

\section{Analysis}
\label{sec:ana}

The optical-UV spectroscopic analysis based on the QMS formalism allowed us to take advantage of the spectral fitting of the emission lines. Each of these lines provides information about the regions where they are emitted, reflecting the gradient of ionization level from the low to high ionization lines. 
Numerous studies have developed different correlations using observed and derived parameters, which are described in this section and will serve as the basis for our analysis. We also consider the FIR-Xr nature of the LS sample.

Regarding previous spectral observations in the LH field, \citet{Lehmann2000} reported 43 quasar spectra in the LH observed between 1996-1998; 12 of them coincide with those of our sample. However, comparing their estimations with ours was impossible because the fitting methodology is different, and they do not consider all the emission lines used in our analysis.

\subsection{Lockman-SpReSO Type 1 AGN in the Quasar Main Sequence context}
\label{sec:LHinQMS}

The QMS in the optical bands has been very useful as it traces accretion-rate-dependent trends. Population A objects show 
higher accretion rates, lower \mbh\ and mostly radio-quiet. A classification in sub-bins has been proposed to group objects with similar spectra and, thus, physical conditions. Populations A1, A2, A3, A4 are defined in terms of increasing \feii\ emission in bins of $\Delta$\rfe\ = 0.5. Objects in A3-A4 sub-bins are identified as 
extreme accretors (xA quasars) in the sense that their accretion rate is close to the Eddington limit. Population B objects 
have low accretion rates, much higher \mbh\, and are mostly radio-loud quasars. Populations in the sub bins B1, B1+, B1++ are defined in terms of increasing $\Delta$FWHM = 4000 \kms\ \citep{marziani10}.

In the ultraviolet range, computing the half-height centroid \cmed\ of \civ\ (see section 5.3, eq. \ref{eq:cmed}) is not a straightforward task for large samples, as the line should be cleaned first from the extra emission, mostly the \civ\ narrow component and \heii. Therefore, we considered the work of \cite{MS14}, who showed that using line ratios in the UV, could be useful to detect highly accreting objects \citep[see also][Jenaro-Ballesteros BSc thesis\footnote{\url{http://132.248.9.195/ptd2023/octubre/0847307/Index.html}}]{martinez-aldamaetal18,Garnica2022,Buendia-Rios2023}. The UV diagram involves line ratios of the intermediate ionization lines of the 1900 \AA\ blend, \ciiio/\aliiio, and \aliiio/\siiiio. This UV diagram also allows the identification of objects with high (\ciiio/\aliiio\ $<$ 1) and low (\ciiio/\aliiio\ $>$ 1) accretion rates, as well as xA objects (\ciiio/\aliiio\ $<$ 1, and \ciiio/\aliiio\ $>$ 0.5). 

\begin{figure}
    \centering
    \includegraphics[width=0.99\linewidth]{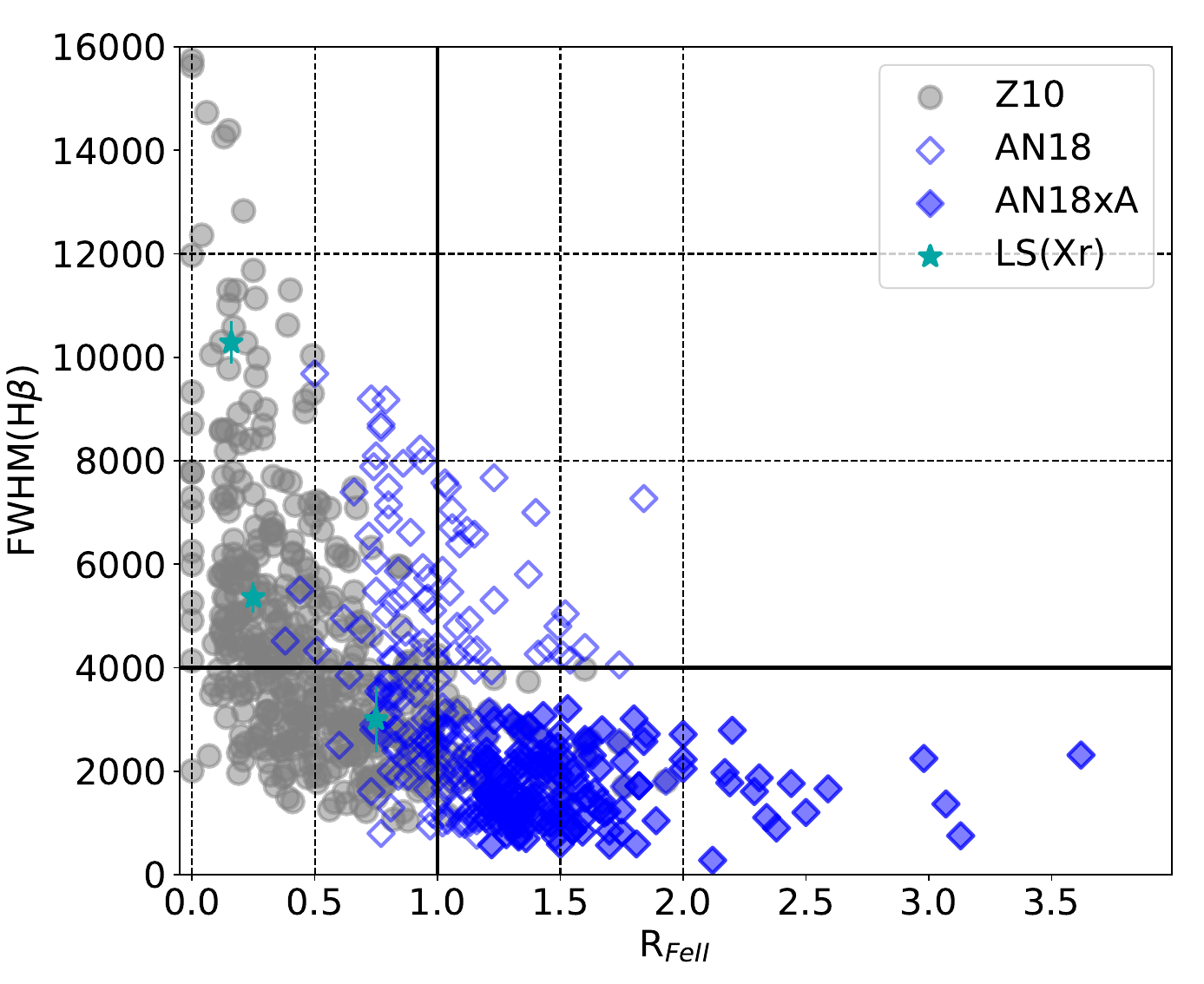}
    \includegraphics[width=0.99\linewidth]{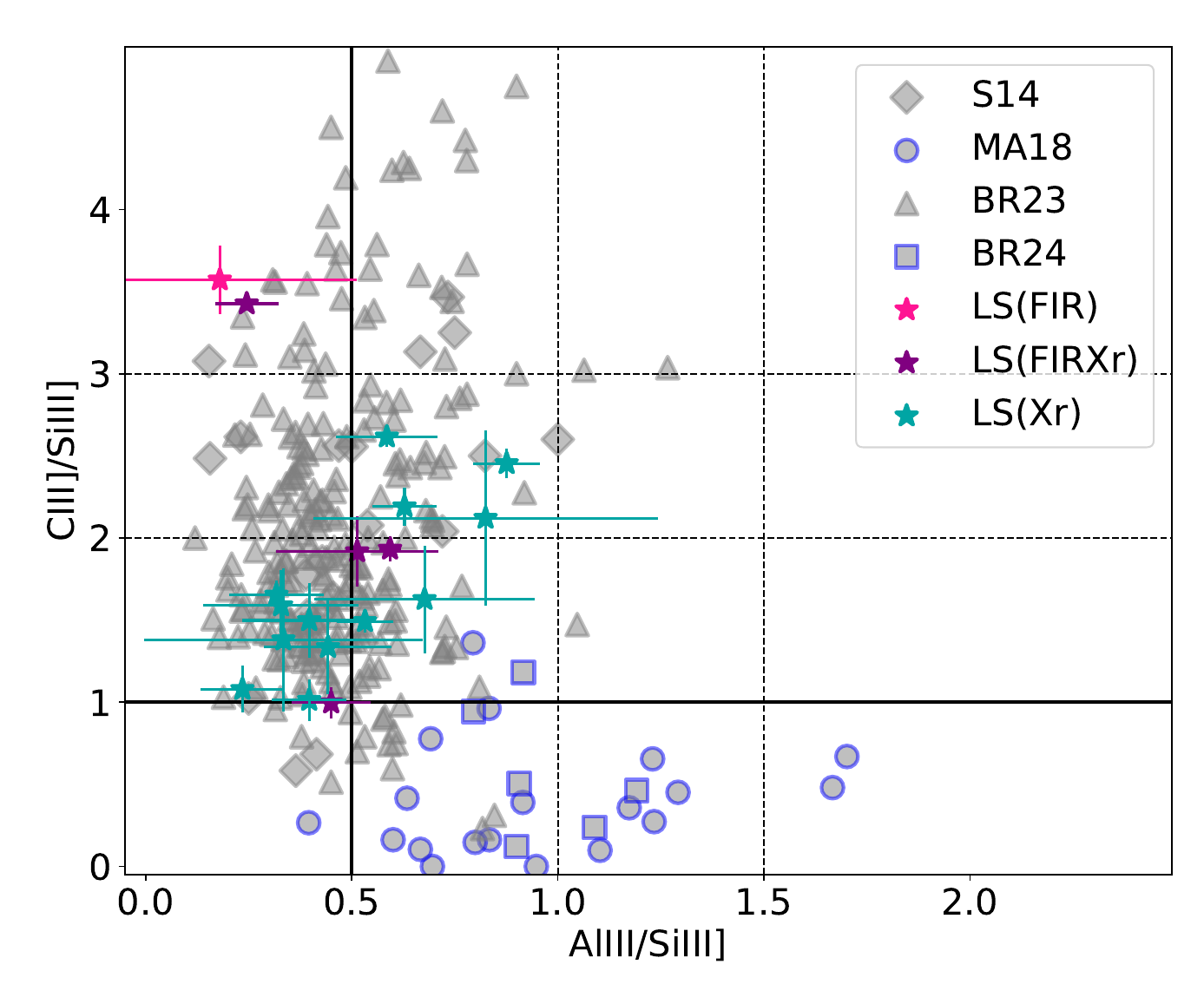}
    \includegraphics[width=1.05\linewidth]{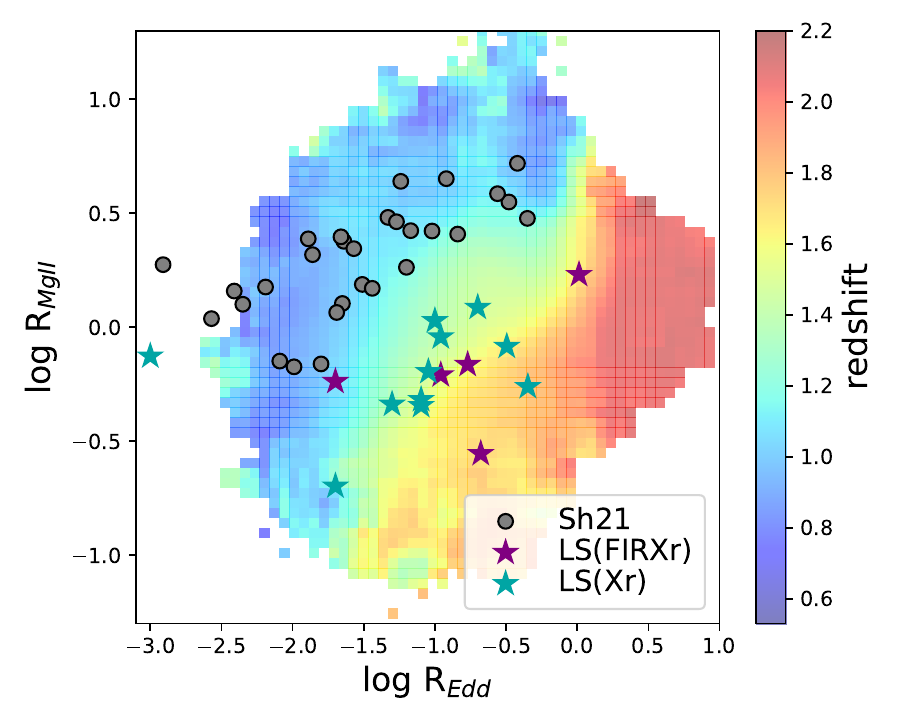}
    \caption{
    QMS in the optical \textit{(Upper panel)} and UV \textit{Middle panel}. The vertical dashed lines are the boundary for xA objects. The horizontal dashed lines are the boundary separation of Pops. A and B. 
    \textit{Lower panel:} Relation between \rmgii\ and \redd. The meanings of the symbols are described in the inner boxes. 
    \bf{Grey circles are low $z$ data from \citet{Shin2021}, and the color code background is the redshift distribution of the SDSS DR17 sample taken from \citet{WuShen2022}}.
    }
    \label{fig:MSs}
\end{figure}

Using our spectral measurements, we reproduced the optical QMS for the three objects at $z\,<\,$ 0.8 and compared the results with archival data from \cite{zamfir10} and \cite{negrete18}. The upper panel of Fig.~\ref{fig:MSs} shows the optical QMS with the objects in LS shown as green stars (the three objects are classified as Xr sources). Two belong to Pop.~B and one to the Pop.~A region. None of them were in the xA domain. 

For the UV range, we compiled spectroscopic information from \citet[][2025 submitted]{martinez-aldamaetal18, sulentic14, negrete14,Buendia-Rios2023} to build the \ciii/\siiii\ vs. \aliii/\siiii\ diagram (shown in middle panel of Fig.~\ref{fig:MSs}). In this UV QMS, the 18 LS objects showing the 1900 \AA\ blend fall into Pop. B region, with three objects at the boundary of Pop. A. In this subsample, we have one FIR object that, together with a FIRXr object, falls in the higher part of the \ciii/\siiii\ values. 
Of the remaining nine objects, three have only \mgiio\ emission. Taking into account the analysis of \citet{Marziani13b, marziani13} that uses the FWHM as a tracer of the Population membership, the three objects are Pop. B. 
The other five objects have \civ\ emission, four with Lorentzian profiles (and then are Pop. A objects) and one with a Gaussian profile (a sign of Pop. B object). 
Finally, one spectrum has only \lya\ emission with Lorentzian profile, 
so it could be considered a Pop. A object. In summary, we have 25 Pop. B and 5 Pop. A objects.

\subsection{\feii\ in the UV}
\defcitealias{Shin2021}{Sh21}

The lower panel of Figure \ref{fig:MSs}, 
shows the relationship between \feii\ UV/\mgiio\ ratio (\rmgii) and \redd\ of the 16 objects showing \mgiio\ (5 FIRXr and 11 Xr sources). We compare the values of the objects in our sample, whose $z$-range is between 0.46 and 1.96, 
with two other samples. 
The first is from the work of \citet[][\citetalias{Shin2021}, grey circles]{Shin2021}, who used 29 objects observed by the Hubble Space Telescope (HST) at $z\,<\,$ 0.37 (with a mean value of 0.04). \citetalias{Shin2021} explained the scenario of the close correlation between \rmgii\ and \redd\ as a consequence of metal cooling, leading to an increase in the gas inflow and, consequently, the accretion rate. 
The second sample corresponds to the SDSS DR17 Quasar Catalogue properties of \citet[][shown in color code]{WuShen2022}. From this second sample, we consider spectra with S/N $>$ 5 in the integrated G band, which is the minimum value of our sample, and log\lbol $>$ 40, which is a similar range to our sample. 
In addition, we define a redshift range from 0.53$\,< z\,<\,$ 2.2. This range assures that the spectral region of \feii\ UV and \mgiio\ are visible within the SDSS DR17 spectral range. We obtained a selection of 208,000  objects. 
As reported by \citetalias[][see their Figure 3]{Shin2021}, we cannot find a redshift dependency with the \rmgii\ ratio for the SDSS DR17 quasars. 
However, we find a clear redshift dependency on the \rmgii-\redd\ relation: the latter decreases as the redshift increases. 
This implies a trend in the \redd, where at lower $z$ we find lower values of \redd, and vice versa, at higher $z$ we have higher values of \redd. In other words, there is a family of \rmgii-\redd\ relations dependent on $z$. 
When we compare with our sample, the \rmgii-\redd\ relation is in agreement with the SDSS DR17 quasars within our redshift range, showing a clear linear relation between \rmgii\ and \redd, but with the expected offset from the \citetalias{Shin2021} relationship. 


\subsection{Winds in the Lockman-SpReSo sample?}
\label{sec:winds}

One of the most important properties of Type 1 AGN is the presence of winds arising from the central region, derived by accretion mechanisms \citep[e.g.][]{Elvis2000}. 
In the BLR, these winds are mainly detected in high-ionization lines, such as \civ, reaching velocities up to -2500 \kms\ \citep{richards11, WuShen2022}. 
\citet{richards11} showed a relationship between the EW and the total shift of the \civ\ line profile, which is argued to be driven by accretion parameters. 
It is worth noting that the approximation in the works that considers thousands of spectra is based on automatic fits to the emission lines, 
and the reported blueshift velocities correspond to the total \civ\ emission. 

Our approximation considers the QMS formalism, where particular initial conditions were considered for different populations, assuming 
two different emitting regions. 
(see Sec.~\ref{sec:meth_considerations}). The isolated blueshifted \civ\ component has been reported in numerous previous works \citep[e.g.][Buendia-Rios et al. 2025 submitted, and references therein]{marziani10,sulentic17,martinez-aldamaetal18}. 

\begin{figure}
    \centering    
    \includegraphics[width=0.99\linewidth]{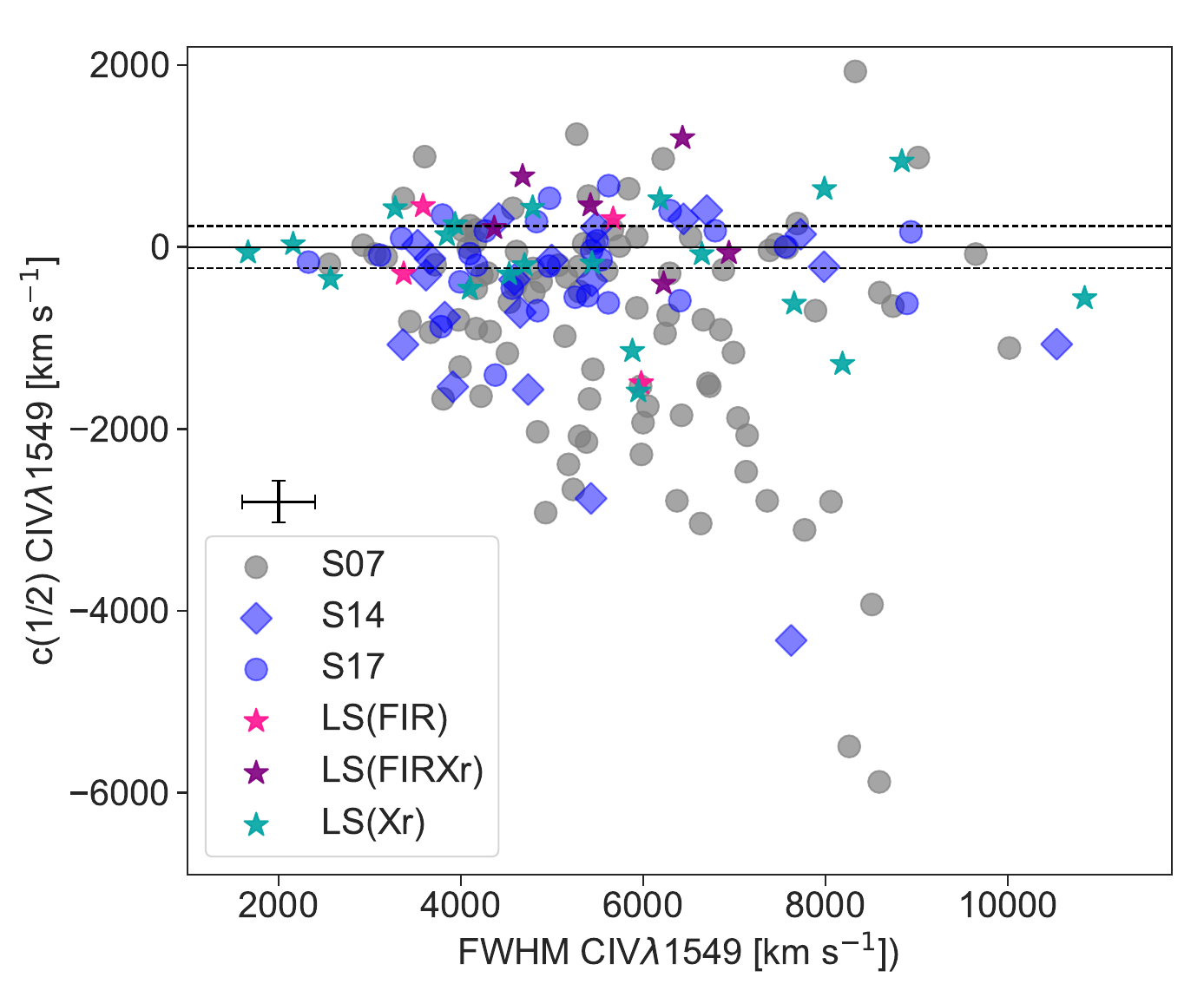}
    \includegraphics[width=0.99\linewidth]{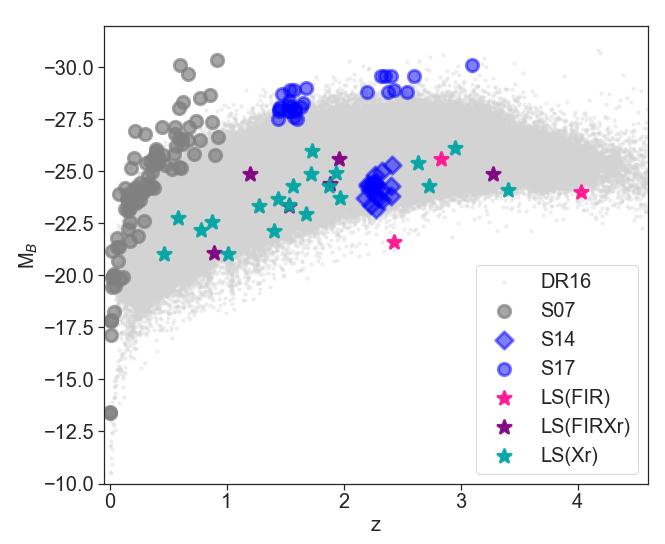}
    \includegraphics[width=0.99\linewidth]{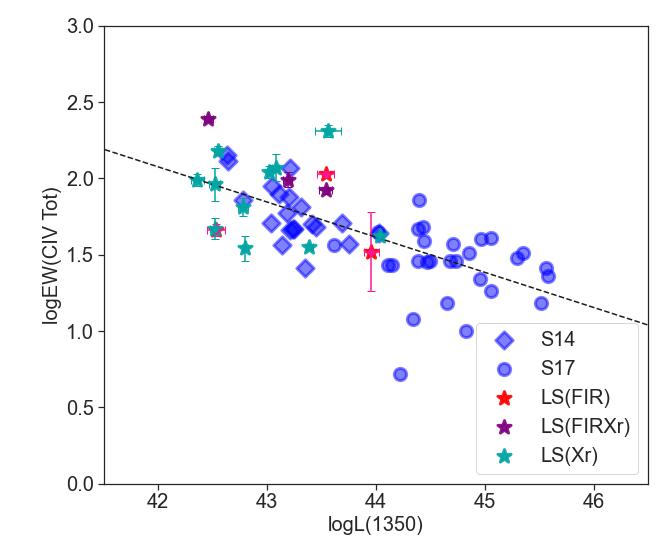}
    \caption{ (Upper Panel) Dependence of the velocity shift at half intensity of \civ, \cmed\ on FWHM(\civ). The solid line is the reference for zero velocity shift. The dashed lines represent the uncertainty range with respect to zero velocity shift. (Middle Panel) Absolute B-magnitude \MB\ vs. redshift $z$. (Lower Panel) Baldwin effect. The dashed line shows the linear fit of our data (Eq. \ref{eq:BE}). The meanings of the symbols are described in the inner boxes.
    }
    \label{fig:CIVshift}
\end{figure}

Another parameter for quantifying the \civ\ blueshift is the half-height centroid c(1/2), defined as 
\begin{equation}
\label{eq:cmed}
    c(1/2) = \frac{v_{r,R}(1/2)+v_{r,B}(1/2)}{2} 
\end{equation}
where $v_{r,R}$ y $v_{r,B}$ are the velocity shifts on the red and blue wings at half-height, respectively \citep{zamfir10}. 
Upper panel of Figure \ref{fig:CIVshift} shows the distribution of \cmed\ versus  FWHM(\civ) from our sample (LS), compared with samples of \citet[][hereafter S07]{sulentic07} for low $z$ objects, and \citet[][hereafter S14, S17]{sulentic14, sulentic17} for high $z$ quasars, for faint and bright samples, respectively. 
In S07 and S17 samples, the tendency to find winds in \civ\ with \cmed\ blueshifts above -4000 \kms\ is noticeable. Winds become less prominent in the S14 and LS samples with blueshifts below -1200 \kms. 
The LS sample has similar $z$ and log\lbol\ ranges as the S14 and S17 samples. However, the S14 sample was selected considering the flux limits of the LS sample at the highest possible $z$. On the other hand, the S17 sample was chosen from the Hamburg ESO \citep[HE;][]{Wisotzki2000} sample as the brightest blue objects at a similar $z$ range to S14. Bluer, brighter quasars at high-$z$ tend to show blueshifts in the HILs.
The middle panel of Figure \ref{fig:CIVshift} compares the $z$-$M_B$ relation from S07, S14, S14, and LS samples, showing that our sample shares a similar $M_B$ magnitude range as the S14 sample, 
including similar spectral properties. For completeness, we also included the absolute magnitudes from the SDSS DR17 data taken from \citet[][, WS22]{WuShen2022}. In both Figures, we do not find any clear tendency with the FIR-Xr classification, except the one with $z$ mentioned in Sec. \ref{sec:selection}.

Finally, in the optical range, 
\oiiill\ is the HIL that traces these winds \citep{zamanov02, negrete18, Grunwald2023}. Optical winds are identified by the presence of a second semi-broad \oiiill\ component shifted to the blue, in addition to the typical narrow component positioned in the restframe. The \oiii\ semi-broad wind component can have FWHM up to 2000 \kms\ and blueshifts up to -2000 \kms\ with respect to the restframe \citep{negrete18}. In quasar spectra, semi-broad \oiiill\ components with offsets larger than -250 \kms\ are considered optical outflows \citep{zamanov02}. 
In our sample, only one object shows a second semi-broad component with an FWHM of 1350 $\pm$ 100 \kms. It has a shift of -25 $\pm$ 27 \kms, therefore, it can not be considered a wind component.


\subsection{Baldwin Effect}
\label{sec:Baldwin}

The anticorrelation between the EW of \civ\ and the continuum luminosity in high-z UV spectra is known as the Baldwin effect \citep[BE,][]{Baldwin1977}. The BE arises from the relationship between the total line flux over the underlying continuum flux and the underlying continuum luminosity, so the correlation is expected. However, as reported in previous works, the slope of the BE relation is not unity, indicating that the line luminosity does not change by the same amount as the underlying continuum does \citep[e.g.,][]{Zheng1993,Wang22}. The origin of the BE relationship has also been discussed in terms of the Eddington ratio. \citet[][see also \citealt{Ge2016}]{bachev04} considered UV spectra within the QMS analysis, finding a systematic decrease of the \civ\ EW as the \redd\ decreases, suggesting that the origin of the \civ\ BE is the \redd\ itself. 

The slope on the BE has been reported to be around -0.24 along a wide range of redshifts. For instance, \citet{Bian2012}, using the SDSS DR7 with 35000 quasars in a redshift range from 1.5 to 5, obtained a slope of -0.238. The lower panel of Figure \ref{fig:CIVshift} shows the BE relation for our sample compared with the high-z samples of S14 and S17. For this analysis, we do not consider the S07 sample as they do not report individual values of the EW(\civ). As discussed in Section \ref{sec:winds}, our sample shares properties similar to the S14 sample and falls in the same region in the BE diagram. We also do not find any tendency with the FIR-Xr classification.
The relationship that we obtain, including S14, S17 and our sample, is:  
\begin{equation}
\label{eq:BE}
    \mathrm{log} \mathrm{EW}(CIV) = (-0.23\pm0.03) \, \mathrm{log} L(1350) + (11.75\pm0.23) 
\end{equation}
with a p-value of -0.67. These values are consistent with the values reported in the literature \citep[e.g.][]{Bian2012}.




\subsection{Comparison with SDSS data}
\label{sec:sdss_comparison}

Given the variable nature of quasars, we searched for the spectra of our sample in the SDSS DR17. 
We found 10 of these with an observation range between 2014-2016, except for object 206641, which was observed in 2002. The observation range coincides with that of the LS observations (between 2014 and 2018; \citetalias{Gonzalez-Otero2023}).
The main difference between SDSS and LS data is that the SDSS spectra have higher resolution but lower S/N. Column 9 of Table \ref{tab:sample} lists the objects found in SDSS DR17. 


We then applied the same criteria and fitting methodology to the spectra observed with our LS sample for the spectral fitting. 
Then, we look for a one-to-one correlation between the results obtained from the SDSS spectra and our LS sample. It is worth noting that these results can be affected mainly by the low S/N of SDSS spectra. The purpose is to look for evidence of variability that may be due to important changes in the physics of the nuclear region.

The line luminosities correlation (L$_{line}$, that considers BC line luminosities of \lya, \siivo, \civo, \ciiio, \aliiio, and \mgiio) is close to the one-to-one relation 
with an offset of -0.20$\pm$0.24. Therefore, we found a systematic shift between SDSS and LS values of 0.2 dex. However, once the offset error is considered, the SDSS values can be recovered within 1$\sigma$ of error. In the case of \lbol\ (in the continuum windows 1350 \AA, 1700 \AA, and 3000 \AA) and FWHM of the BCs of \lya, \siivo, \civo, \ciiio, \aliiio, and \mgiio) comparisons, 
the offsets are -0.04 $\pm$ 0.28 and -50 $\pm$ 820, respectively. Considering the errors, we can conclude that L$_{line}$, \lbol, and the FWHM from the SDSS and LS values are consistent. In addition, we did not find indications of variability between the two samples once scatter was considered.

\subsection{Extinction Analysis}
\label{sec:extinction}

We examine the effects of extinction in our Galaxy on the flux estimations of quasars. For this purpose, we compare the colors $g_{AB}-r_{AB}$ of our quasar LS sample (located around 57.5$^{\circ}$ in declination) with other quasar samples taken from the SDSS DR17 in the galactic plane (between $\pm$ 15$^{\circ}$ in declination, group Gal Plane), and the galactic north pole (above 60$^{\circ}$ in declination, group North Pole). We restrict the three quasar sample to a redshift around $z$ = 1-2 since most of our LS objects are in this range (see Fig. \ref{fig:zHisto}).
The average color $g_{AB}-r_{AB}$ for the corrected and non-corrected extinction \citep[following the][extinction law]{Cardelli1989_CCM} of the three groups is:

\begin{itemize}
    \item LS: \,\,\,\,\,\, 0.233 $\pm$ 0.299; dered 0.213 $\pm$ 0.276
    \item Gal Plane: 1.145$\pm$0.723; dered 0.200$\pm$0.358
    \item North Pole: 0.195$\pm$0.228; dered 0.174$\pm$0.227
\end{itemize}

In conclusion, the current extinction corrections are reliable, especially for objects in the galactic plane, since we can recover the color values.

\section{Summary}
\label{sec:summ}

This paper describes the characteristics of Type 1 AGN identified by the Lockman-SpReSO project. We present high-quality spectra of 30 quasars with a redshift range of 0.462 to 4.967. A detailed spectral fitting method based on the QMS phenomenology prescription was applied to deconvolve the \hb, \mgiio, 1900\AA\ blend, \civ, \siiv, and \lya\ regions found in the spectra of the sample. The overall characteristics of the sample are as follows: 
\begin{itemize}
    \item log\lbol: 44.85 -- 47.87
    \item log\mbh: 7.59 -- 9.80
    \item log\redd: $-$1.70 -- 0.56
    \item \MB:\,\,\,\,\, $-$20.46 -- $-$26.14
\end{itemize}

The detailed optical-UV spectral splitting allows us to perform different analyses based on correlations of the emission line parameters, properties, and physical conditions of the emission regions reported in previous studies, as described below.

\begin{itemize}
    \item We found a wide spectral diversity in our sample from the QMS analysis using optical and UV emission lines. The low-z sample (three objects) falls in Populations A2, B1, and B1+, while quasars of the high-z sample, 23 are Pop. B and 4 Pop. A objects. None of the LS quasars was found in the xA domain. 
    \item We reproduced the relationship between the Eddington ratio \redd, the fundamental parameter underlying the spectral differences in AGN and the FWHM of the virial components, with an anticorrelation described by log \redd\ = -0.17 FWHM$_{1000}$(BC) + 0.27  with a correlation coefficient $r_p$ = -0.7, similar to that reported by \citet{marziani01}.
    \item Using the measurements of the spectra showing \mgii\ (in the range 0.462 $> z >$ 1.956), we recovered the relation \redd-\rmgii, consistent with the results of \citet[][for objects with z$<$0.37]{Shin2021} and the SDSS DR17 quasar catalog sample in a $z$ range similar to the LS. 
    \item We also looked for signs of winds in spectra showing the HIL \civ\ in the UV. We computed its half-height centroid \cmed\ as a proxy for the blueshift of the line, finding values of the \cmed\ between 941 and -1587 \kms. Compared with the other low- and high-$z$ samples of S07, S14, and S17, the LS quasars follow the trends of the S14 sample, which shares a similar range of $z$ and absolute magnitudes.
    \item We also consider the anti-correlation of the Baldwin effect. We find a trend in terms of log L(1350) vs. log \civo(EW), with a slope of -0.23 $\pm$ 0.03 dex, similar to that reported in the literature, especially by \citet{Bian2012} who used the SDSS DR 16 quasar catalog sample. 
    \item We find 12 spectra of our sample in the SDSS DR17 archive, for which we use the same spectral fitting methodology as with the LS spectra. The principal difference between the SDSS and LS spectra is the low S/N ratio of the former. We found no evidence of spectral variability when comparing values of line luminosities, FWHM, and bolometric luminosities. 
    \item Finally, we perform an extinction analysis comparing the $g_{AB}-r_{AB}$ colors of our sample with an SDSS DR17 quasar sample. The effects of extinction in the LS field are similar to the ones in the Galactic North Pole, while using the extinction law of \citet{Cardelli1989_CCM} allows recovery of dereddened colors for objects in the Galactic Plane.
\end{itemize}
It is worth noticing that despite the LS sample was selected for the large emission in the FIR using Herschel data at 100 and 160 $\mu m$, and X-ray data; we found no evidence of different behavior from those AGN in non-obscured or with different criteria selected samples, such as the SDSS. 
    
\section*{Acknowledgements}

CAN and HJIM thank the support of the CONAHCyT projects 2022-320020, CBF2023-2024-1418, and the DGAPA-UNAM grants IA104325 and IN111422. HJIM thanks support from CONAHCyT project CF-2023-G-543. ICG and EB acknowledge financial support from DGAPA-UNAM grant IN-119123 and CONAHCYT grant CF-2023-G-100. MHE acknowledges support from CONAHCYT program Estancias Posdoctorales por México. MC acknowledges funds by grant PID2022-136598NB-C33 funded by MCIN/AEI/10.13039/501100011033 and by “ERDF A way of making Europe”. TM, MEC, and MHE thank the support from UNAM DGAPA PAPIIT IN 114423. HMHT acknowledges support from grants CF-G-543 CONAHCYT and CF-2023-G-1052 CONAHCYT. MSP acknowledges the support of the Spanish Ministry of Science, Innovation and Universities through the project PID--2021--122544NB—C43.
This work was supported by the Evolution of Galaxies project, of reference PID2021-122544NB-C41 within the Programa estatal de fomento de la investigaci\'on cient\'ifica y t\'ecnica de excelencia del Plan Estatal de Investigaci\'on Cient\'ifica y T\'ecnica y de Innovaci\'on of the Spanish Ministry of Science and Innovation/State Agency of Research MCIN/AEI. This article is based on observations made with the Gran Telescopio Canarias at Roque de los Muchachos Observatory on the island of La Palma. 
Funding for the Sloan Digital Sky Survey IV has been provided by the Alfred P. Sloan Foundation, the Heising-Simons Foundation, the National Science Foundation, and the Participating Institutions. SDSS acknowledges support and resources from the Center for High-Performance Computing at the University of Utah. SDSS telescopes are located at Apache Point Observatory, funded by the Astrophysical Research Consortium and operated by New Mexico State University, and at Las Campanas Observatory, operated by the Carnegie Institution for Science. The SDSS web site is \url{www.sdss.org}. SDSS is managed by the Astrophysical Research Consortium for the Participating Institutions of the SDSS Collaboration, including Caltech, The Carnegie Institution for Science, Chilean National Time Allocation Committee (CNTAC) ratified researchers, The Flatiron Institute, the Gotham Participation Group, Harvard University, Heidelberg University, The Johns Hopkins University, L'Ecole polytechnique f\'{e}d\'{e}rale de Lausanne (EPFL), Leibniz-Institut f\"{u}r Astrophysik Potsdam (AIP), Max-Planck-Institut f\"{u}r Astronomie (MPIA Heidelberg), Max-Planck-Institut f\"{u}r Extraterrestrische Physik (MPE), Nanjing University, National Astronomical Observatories of China (NAOC), New Mexico State University, The Ohio State University, Pennsylvania State University, Smithsonian Astrophysical Observatory, Space Telescope Science Institute (STScI), the Stellar Astrophysics Participation Group, Universidad Nacional Aut\'{o}noma de M\'{e}xico, University of Arizona, University of Colorado Boulder, University of Illinois at Urbana-Champaign, University of Toronto, University of Utah, University of Virginia, Yale University, and Yunnan University.



\bibliography{references}{}
\bibliographystyle{aasjournal}

\end{document}